\pdfoutput=1
\documentclass[acmsmall,screen,nonacm]{acmart}
\usepackage{amsmath}
\usepackage{stmaryrd}
\usepackage{enumitem}
\usepackage{pgfplots}
\usepackage{subcaption}
\usepackage{mathtools}
\usepackage{hyperref}
\usepackage{cleveref}
\usepackage{tikz}
\usepackage{xcolor}
\usepackage{newverbs}
\usepackage{fancyvrb}
\usetikzlibrary{positioning, arrows.meta, calc, bending, backgrounds, shapes.misc, shapes.geometric}
\pgfplotsset{compat=1.18}
\usepgfplotslibrary{fillbetween}

\setcopyright{none}
\renewcommand\footnotetextcopyrightpermission[1]{}
\newverbcommand{\errorverb}
  {\begin{lrbox}{\verbbox}}
  {\end{lrbox}\colorbox{red!30}{\box\verbbox}}

\AtBeginDocument{%
  }

\acmYear{2026}
\acmJournal{JACM}

\begin{document}

\title{\textsc{AGDecode}: Semantic Attribute Grammar Derivatives for Safety-Constrained Decoding of Large Language Models}

\author{Rajan Maghera}
\affiliation{%
  \institution{University of Toronto}
  \department{Department of Computer Science}
  \city{Toronto}
  \country{Canada}
}
\orcid{0009-0003-3947-2865}
\email{maghera@cs.toronto.edu}
\author{Asher Tempest Arya}
\affiliation{%
  \institution{University of Toronto}
  \department{Department of Computer Science}
  \city{Toronto}
  \country{Canada}
}
\orcid{0009-0008-0774-7613}
\email{asher.arya@mail.utoronto.ca}
\author{Aren A. Babikian}
\affiliation{%
  \institution{McGill University}
  \department{Department of Electrical and Computer Engineering}
  \city{Montreal}
  \country{Canada}
}
\orcid{0000-0002-8108-0043}
\email{aren.babikian@mcgill.ca}
\author{Marsha Chechik}
\affiliation{%
  \institution{University of Toronto}
  \department{Department of Computer Science}
  \city{Toronto}
  \country{Canada}
}
\orcid{0000-0002-6301-3517}
\email{chechik@cs.toronto.edu}
\author{Maryam Mehri Dehnavi}
\affiliation{%
  \institution{University of Toronto}
  \department{Department of Computer Science}
  \city{Toronto}
  \country{Canada}
}
\additionalaffiliation{%
  \institution{NVIDIA Research}
}
\orcid{0000-0002-2719-8788}
\email{mmehride@cs.toronto.edu}

\theoremstyle{acmdefinition}
\newtheorem{axiom}{Axiom}

\begin{abstract}

    Large language models (LLMs) are increasingly used for code generation, motivating techniques that guarantee the structural and semantic correctness of generated programs. Constrained decoding enforces correctness during generation by restricting token choices to those that admit at least one valid completion. While existing approaches can enforce syntactic constraints and selected semantic properties such as type safety, they either target specific properties or rely on operational pruning mechanisms that are difficult to generalize to data-dependent, inductive program properties. We present \textsc{AGDecode}, a programmable framework for semantic constrained decoding based on least-fixed-point semantics. \textsc{AGDecode} extends attribute grammars to partially constructed syntax trees, and reduces semantic constrained decoding to a realizability problem over such trees. The framework compiles semantic specifications into systems of relational constraints that are discharged by external verification oracles, such as SMT and CHC solvers, cleanly separating language semantics from constraint-solving machinery. We prove that our approach is sound and complete relative to an ideal verification oracle. In practice, solvers inevitably encounter uncertain queries; conservatively, our system can reject tokens for soundness or speculatively continue to generate until later queries resolve the entire generation. We mechanize the core metatheory in Rocq and evaluate the framework on semantic-constrained generation of C programs, enforcing path-sensitive safety properties including null-pointer safety, division-by-zero freedom, and initialized-variable usage. Our results demonstrate that solver-backed semantic constrained decoding can enforce rich semantic guarantees while incurring modest computational overhead at regular program boundaries.

\end{abstract}

\maketitle

\section{Introduction}

Recent advances in large language models (LLMs) have significantly expanded their applications in structured domains, most notably automated program generation. However, LLMs provide no native guarantees that their generated text conforms to structural or semantic invariants.

To mitigate this limitation,  \emph{constrained decoding} has emerged as a powerful paradigm for enforcing safety properties during generation~\cite{scholak_picard_2021, poesia_synchromesh_2022, beurer-kellner_guiding_2024, willard_efficient_2023}. Constrained decoding ensures that the final output is correct-by-construction by restricting the decoder's token selection to continuations that admit at least one completion satisfying a given specification. While syntactic enforcement via context-free grammars is well-explored~\cite{geng_grammar_2023, park_grammar-aligned_2025, ugare_syncode_2024}, syntax alone is insufficient for guaranteeing deeper properties. \emph{Semantic constrained decoding} addresses this gap by enforcing static properties such as type safety and variable scoping \cite{mundlerTypeConstrainedCodeGeneration2025, nagyChopChopProgrammableFramework2025}.

Early attempts to enforce semantic constraints have largely relied on ad-hoc or single-property encodings. Most notably,
\citet{mundlerTypeConstrainedCodeGeneration2025}
introduce a decoder tailored explicitly for type safety by evaluating type-checking logic over a program's incremental string representation. However, monolithic approaches tied to a single semantic property scale poorly to generalized constraints.
Evaluating semantic properties directly on a generated prefix
requires the decoder to look ahead and reason about every
permissible decoder token continuation.

\paragraph{Motivating example}
Consider the task of enforcing selected undefined-behaviour (UB) safety properties on a C program while an LLM generates the program. Consider an incomplete generation:
\begin{center}
\verb|void foo(int num) { int *x = malloc(sizeof(int)); |\errorverb|*x| \verb| = 16 / | \errorverb|num;|
\end{center}
This incomplete generation contains two potential UB violations, highlighted in red, that cannot be rectified by generating more of the program. The first UB violation is a null-pointer dereference if \verb|x == NULL|, and the second is a divide-by-zero violation if \verb|num == 0|. Semantic constrained decoding must therefore permit the dereference of  \verb|x| and the use of \verb|num| as a divisor only when the required non-nullness and non-zero facts can be established.

\paragraph{Programmable semantic constrained decoding frameworks} To provide a generalized solution to semantic constrained decoding, \citet{nagyChopChopProgrammableFramework2025} introduce \textsc{ChopChop}, the first programmable framework for semantic constrained decoding. \textsc{ChopChop} addresses the problem of expressivity and programmability of constrained decoders by allowing for a programmable set of semantics. The space of all possible abstract syntax tree (AST) continuations for a given prefix is represented compactly as regular codata, reducing constraint enforcement to a realizability problem solved via coinduction. A user specifies a language grammar and a system of corecursive \emph{semantic pruner} functions to eliminate invalid ASTs from the program space. While highly expressive, managing program spaces via coinductive coterms and operational pruners introduces fundamental challenges for general program verification:

(1) \textbf{Representational overhead in non-regular domains}. Courcelle's theorems~\cite{courcelle_infinite_trees_1983} show that coterms are highly effective for capturing infinite control-flow behaviours or behavioural bisimulations. However, they are ill-suited for modelling data theories, such as bit-precise arithmetic or path conditions. Forcing these complex data states into a structural tree requires a large number of unique symbols, resulting in a state-space explosion.

(2) \textbf{Coupling of semantics and search}. Because \textsc{ChopChop}'s pruners are formulated as operational, total functions over a trace space, they simultaneously define the language's semantics and the execution oracle. Consequently, these pruners operate as algorithmic black boxes, preventing the underlying framework from performing global logical optimizations.

Many safety properties are inductive. For example, assertion reachability and memory safety are defined over the set of reachable program states, a \emph{least fixed point} capturing the variable-value tracking required in the motivating example. Traditional program verification uses this structure by reducing a program to a set of relational verification conditions (VCs) and discharging them to efficient Satisfiability Modulo Theories (SMT) and Constrained Horn Clause (CHC) solvers \cite{de_moura_z3_2008, beklemishev_horn_2015}. Proving safety then reduces to a reachability query over that least fixed point, which is the kind of query these solvers are designed to decide. \textsc{ChopChop} instead represents the space of completions as coinductive coterms, whose semantics is a \emph{greatest fixed point}. This fits inductive properties poorly. A data-dependent inductive invariant has no compact coterm representation, because distinguishing states by their bit-precise values requires a separate structural symbol for each valuation and causes a state-space explosion.

In this paper, we ask:
\begin{center}
\emph{How can we design a programmable framework that leverages inductive least-fixed-point semantics to integrate automated verification oracles into constrained decoding?}
\end{center}

\paragraph{Our approach} To overcome these limitations, we formalize constraint
evaluation using Attribute Grammars (AGs)~\cite{knuth_semantics_1968}. While traditional
AGs evaluate semantic equations over complete syntax trees, constrained decoding
demands reasoning over an incomplete token prefix spanning an infinite space of
possible completions. Our framework, \textsc{AGDecode}, addresses this challenge via
a relational AG formulation that introduces placeholder \emph{holes} at ungenerated
syntax nodes. Rather than executing an operational traversal over a partial tree,
\textsc{AGDecode} translates the structural grammar rules into a uniform system of
relational constraints and defers the semantic realizability query to an external
oracle (e.g., SMT solver) capable of solving recursive rules efficiently.
This architecture cleanly decouples the language's semantic specification
from its checking machinery. We prove that our framework is sound and complete with respect to ideal evaluation. In practice, oracles may emit uncertain results. Thus, we show in evaluation that speculatively-generated decoder tokens can be verified in retrospect to ensure soundness.

This paper makes the following contributions:
\begin{enumerate}[label=(\arabic*)]
    \item a novel programmable framework for semantic constrained decoding based on least-fixed-point semantics (Sec.~\ref{sec:overview});
    \item a formalization for residual attribute grammars used to represent partially-constructed syntax trees (Sec.~\ref{sec:semantic_ags});
    \item a formal connection tying semantic constrained decoding to generating the partially-constructed syntax trees (Sec.~\ref{sec:semantic_derivatives});
    \item
  an embedding to compile residual attribute grammar semantics into systems of CHCs. (Sec.~\ref{sec:symbolic_encoding}); and
  \item   an evaluation of our framework showcasing the generation of semantically verified programs while incurring small computational overhead (Sec.~\ref{sec:evaluation}).
\end{enumerate}
A mechanization of the core meta-theorems of the framework in the Rocq proof assistant is provided as supplementary material\footnote{{Proof scripts: \href{https://doi.org/10.5281/zenodo.21287418}{https://doi.org/10.5281/zenodo.21287418}}}.

\section{Background}
\label{sec:background}

\subsection{Semantic Constrained Decoding}
Constrained decoding is implemented by a monitor, a procedure that tracks the decoder tokens generated so far and determines which decoder tokens are permitted to follow \cite{agrawal_monitor-guided_2023, nagyChopChopProgrammableFramework2025}. A monitor maintains a state summarizing the generated prefix and exposes an acceptance predicate that, for a candidate next decoder token, reports whether appending it keeps at least one valid completion reachable. At each decoding step the monitor is queried across the vocabulary and the model's next-token distribution is masked to the accepted set. Consuming a decoder token then advances the monitor's state.
The parsing with derivatives approach handles the partial parse tree formed by a continuous token stream \cite{might_parsing_2011, brzozowski_derivatives_1964}.
The derivative of a formal language $\mathcal{L}$ with respect to a terminal symbol $c$ is defined as the set of all strings $w$ such that the concatenated string $cw$ is in $\mathcal{L}$ \cite{brzozowski_derivatives_1964}. Formally:
    $\partial_c(\mathcal{L}) = \{ w \mid cw \in \mathcal{L} \}$.
In other words, the derivative is the set of strings that are valid continuations of the prefix $c$.
In derivative parsing, taking the derivative of a grammar with respect to a character effectively advances the parser's state.
Rather than rebuilding a parse tree from scratch for every new character, the parse tree is modified using a specialized grammar synthesized from the derivative operator. This grammar describes all valid parse trees that can follow the consumed prefix.

\paragraph{Parsing Derivatives for Constrained Decoding}

Derivative parsing can be applied to constrained decoding by treating the character stream as successive derivatives of a parse tree. In the case that the derivative would leave the language empty, the last consumed character is invalid. Derivative parsing realizes a syntactic monitor. Its state is the active grammar, and taking the derivative with respect to a character advances that state. A decoder token is accepted when the resulting derivative is nonempty: $\mathsf{accept}(\mathcal{L}, c) \equiv{} \partial_c(\mathcal{L}) \neq \emptyset{}$.
The predicate holds exactly when some string in $\mathcal{L}$ continues the consumed prefix through $c$. A syntactic monitor constrains the shape of the output and nothing further. Semantic constrained decoding keeps this interface and strengthens the acceptance predicate, admitting a token only when a completion exists that also satisfies a semantic specification.

\subsection{Attribute Grammars}
AGs can be used to connect the syntax of a program and its semantic properties. Originally introduced in \citet{knuth_semantics_1968}, AGs extend context-free grammars (CFGs) by associating declarative attributes with grammar symbols, such as the type of an expression or the set of variables in scope at a node, and providing semantic rules to compute these values over the parse tree \cite{aho_compilers_2006}. In this section, we introduce the core mechanics of attribute grammars through a working example.

\subsubsection{Synthesized attributes}
Consider a minimal language consisting of literal values (integers, floats, strings) and a binary sum operation. A pure CFG dictates the shape of valid expression trees, but it cannot express the context-sensitive constraint that addition is only well-defined for compatible types, nor can it dictate how those types promote.
We define the syntax of our example expression language as follows:
$    E \Coloneqq E_1 + E_2 \mid \text{int\_const} \mid \text{float\_const} \mid \text{string\_const}$.
To enforce typing rules, we decorate the non-terminal $E$ with a \emph{synthesized} attribute, $E.\tau$, which represents the type of the evaluated expression ($\tau \in \{\mathsf{Int}, \mathsf{Float}, \mathsf{String}\}$). Because the type of a parent node in our language depends exclusively on the attributes of its immediate children, this is an instance of an \emph{S-attributed grammar} \cite{aho_compilers_2006}. In an S-attributed system, all attributes are synthesized, meaning they can be evaluated strictly bottom-up via a post-order traversal of the parse tree.

We model the type determination by attaching semantic rules to each production. For the base cases (the literals), the type attribute is instantiated directly from the lexemes:
\begin{align*}
    E &\to \text{int\_const} & & E.\tau = \mathsf{Int}\\
    E &\to \text{float\_const} & & E.\tau = \mathsf{Float}\\
    E &\to \text{string\_const} & & E.\tau = \mathsf{String}
\end{align*}

For the recursive addition production, the semantic rule computes the resulting \emph{promoted} type. In our case, integers get promoted to floats, and strings are not supported at all (i.e., no concatenation).
We introduce a partial promotion function, $\mathsf{promote}(\tau_1, \tau_2)$, which maps two operand types to a resulting type.
\begin{align*}
    E &\to E_1 + E_2 & & E.\tau = \mathsf{promote}(E_1.\tau, E_2.\tau)
\end{align*}
where the promotion relation is explicitly defined as:
\begin{align*}
    \mathsf{promote}(\mathsf{Int}, \mathsf{Int}) &= \mathsf{Int}
    &\mathsf{promote}(\mathsf{Float}, \mathsf{Float}) &= \mathsf{Float} \\
    \mathsf{promote}(\mathsf{Int}, \mathsf{Float}) &= \mathsf{Float}
    &\mathsf{promote}(\mathsf{Float}, \mathsf{Int}) &= \mathsf{Float} \\
    \mathsf{promote}(\_, \_) &= \mathsf{Invalid}
\end{align*}

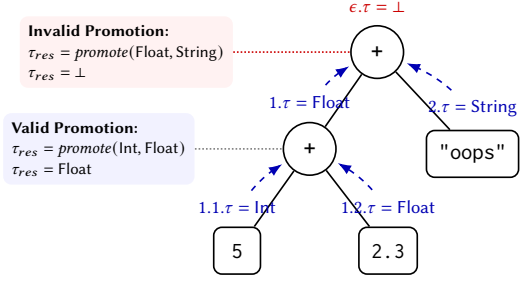
\begin{figure}[t]
\centering
\resizebox{0.5\textwidth}{!}{
\begin{tikzpicture}[
    >=Latex,
    node distance=2.5cm and 3.5cm,
    ast/.style={circle, draw, thick, minimum size=9mm, font=\Large\bfseries},
    leaf/.style={rectangle, draw, thick, minimum size=8mm, rounded corners, font=\ttfamily\large, inner sep=6pt},
    attr/.style={text=blue!70!black, font=\small\sffamily, align=center},
    flow/.style={->, thick, blue!70!black, dashed, shorten >=2pt, shorten <=2pt},
    edge_label/.style={fill=white, inner sep=2pt, font=\footnotesize\sffamily, text=blue!70!black},
    eval_box/.style={rectangle, rounded corners, fill=white, font=\footnotesize\sffamily, align=left, inner sep=4pt}
]

\node[ast] (root) {+};
\node[ast, below left=1cm and 0.5cm of root] (plus) {+};
\node[leaf, below right=1cm and 0.5cm of root] (str) {"oops"};

\node[leaf, below left=1cm and 0.5cm of plus] (five) {5};
\node[leaf, below right=1cm and 0.5cm of plus] (float) {2.3};

\draw[thick] (root) -- (plus);
\draw[thick] (root) -- (str);
\draw[thick] (plus) -- (five);
\draw[thick] (plus) -- (float);

\node[attr, above=0.1cm of five] (t_five) {$1.1.\tau = \mathsf{Int}$};
\node[attr, above=0.1cm of float] (t_float) {$1.2.\tau = \mathsf{Float}$};
\node[attr, above=0.1cm of str] (t_str) {$2.\tau = \mathsf{String}$};
\node[attr, above=0.1cm of plus] (t_plus) {$1.\tau = \mathsf{Float}$};
\node[attr, above=0.1cm of root, text=red!80!black] (t_root) {$\epsilon.\tau = \bot$};

\draw[flow] (t_five) to[bend left=10] (plus);
\draw[flow] (t_plus) to[bend left=10] (root);

\draw[flow] (t_float) to[bend right=10] (plus);
\draw[flow] (t_str) to[bend right=10] (root);

\node[eval_box, left=1.0cm of plus, xshift=-0.5cm, fill=blue!5] (eval_plus) {
    \textbf{Valid Promotion:}\\
    $\tau_{res} = \mathit{promote}(\mathsf{Int}, \mathsf{Float})$\\
    $\tau_{res} = \mathsf{Float}$
};
\draw[gray, densely dotted, thick] (eval_plus.east) -- (plus.west);

\node[eval_box, left=2.0cm of root,  fill=red!5] (eval_root) {
    \textbf{Invalid Promotion:}\\
    $\tau_{res} = \mathit{promote}(\mathsf{Float}, \mathsf{String})$\\
    $\tau_{res} = \bot$
};
\draw[red!80!black, densely dotted, thick] (eval_root.east) -- (root.west);

\end{tikzpicture}}
\caption{Visualization of S-attributed evaluation over the parse tree for \texttt{(5 + 2.3) + "oops"}.
         The structural parse tree is connected via solid lines, while the dashed outer edges trace the bottom-up synthesis of type attributes ($E.\tau$). At the root node, the evaluation relies on a partial $\mathit{promote}$ function which maps to $\bot$, explicitly violating the grammar's semantic constraints and rejecting the program space.}
\label{fig:type_flow}
\label{fig:ag_example}
\end{figure}

You may expect that the premise for \textsc{Sum} includes a proposition that the type is valid (i.e., $\tau_{res} \neq \mathsf{Invalid}$). Indeed, omitting attribute definitions on invalid branches does not constrain invalid behaviour. Since attribute rules must be completely specified for every production, all paths upward must explicitly define how an attribute is transferred, even if it means propagating an explicit $\mathsf{Invalid}$ when a promotion is disallowed. Attributes are evaluated with the syntax $\llbracket t \rrbracket \downarrow \pi.a$ where $t$ is a tree, $\pi \in \mathbb{N}^*$ is the index of the tree and $a$ is the attribute. Figure~\ref{fig:ag_example} visualizes this flow of information through a parse tree when evaluating $\llbracket t \rrbracket \downarrow \epsilon.\tau$ for the expression \texttt{(5 + 2.3) + "oops"}.

\subsubsection{Inherited attributes}
While S-attributed grammars elegantly model bottom-up data flow, they lack the capacity to push contextual requirements downward into the tree. To illustrate this limitation, we extend our language to include variable declarations with explicit type annotations:
\begin{align*}
    P &\Coloneqq S \mid S \ P_1 \\
    S &\Coloneqq T \ \mathit{id} = E \\
    T &\Coloneqq \text{\texttt{int}} \mid \text{\texttt{float}} \mid \text{\texttt{str}} \\
    E &\Coloneqq E_1 + E_2 \mid \text{int\_const} \mid \text{float\_const} \mid \text{string\_const} \mid \mathit{id}
\end{align*}
Consider the base case for the $\mathit{id}$ literal. The type of a constant is defined statically by the syntax, but the type of a variable depends on its definition, which is fetched from the symbol table, an attribute that belongs to parent nodes.

To resolve this, contextual information must be passed from the parent nodes down to their children. The \emph{inherited} attribute, $E.\eta$, contains the \emph{symbol table} and is dictated by the surrounding context. By pairing the synthesized actual type ($E.\tau$) with the inherited symbol table ($E.\eta$), our model transitions into an \emph{L-attributed grammar}, a specific type of attribute grammar allowing information to flow not only down and up the parse tree, but also from left-to-right for siblings \cite{aho_compilers_2006}.\footnote{The underlying framework does not require any particular order for the flow of information. \textsc{AGDecode} is not limited to L-attributed grammars.}
The top-level declaration rule is formalized through the injection of the symbol table into the expression's inherited attribute. This symbol table is used to perform symbol lookups.
\begin{align*}
    E &\to \mathit{id} & & E.\tau = \mathsf{lookup}(E.\eta, \mathit{id})
\end{align*}
where the lookup relation is explicitly defined as
\begin{align*}
\mathsf{lookup}(\eta, \mathit{id}) = \begin{cases}
\tau & \text{if } (\mathit{id}, \tau) \in \eta \\
\mathsf{Invalid} & \text{otherwise}
\end{cases}
\end{align*}
Rules to synthesize type names to their given types $T.\tau_{decl}$ are analogous to the constant cases.
\begin{align*}
    T &\to \texttt{int} & & T.\tau_{\mathit{decl}}= \mathsf{Int}\\
    T &\to \texttt{float} & & T.\tau_{\mathit{decl}} = \mathsf{Float}\\
    T &\to \texttt{str} & & T.\tau_{\mathit{decl}} = \mathsf{String}
\end{align*}

The symbol table must be queried and updated when a variable is defined only if the variable hasn't been defined before. The attribute grammar gives us control over how attributes are passed and in which direction. In this case, the symbol table is passed down through $E.\eta$. This rule updates the symbol table, so the new symbol table must be passed up as a synthesized attribute:
\begin{align*}
    S &\to T \ \mathit{id} = E & & \
    E.\eta = S.\eta \\
    & & & S.\eta_{\mathit{res}} =  \mathsf{update}(S.\eta, \mathit{id}, E.\tau, T.\tau_{decl})
\end{align*}
where the symbol table update is explicitly defined as
\begin{align*}
    \mathsf{update}(\eta, \mathit{id}, \tau, \tau_{decl}) = \begin{cases}
\eta \cup \{(\mathit{id}, \tau)\} & \text{if } \forall \tau'. \, (\mathit{id}, \tau') \notin \eta, \tau = \tau_{decl} \\
\mathsf{Invalid} & \text{otherwise}
\end{cases}
\end{align*}
Previous semantic rules must be updated to pass the newly inherited expression down. Because the symbol table is never updated through an expression, the synthesis of a new symbol table is not required.
\begin{align*}
    E &\to E_1 + E_2 & & E_1.\eta = E.\eta \\
    &&& E_2.\eta = E.\eta \\
    & & & E.\tau = \mathsf{promote}(E_1.\tau, E_2.\tau)
\end{align*}
Finally, statements are chained via the $P \to S\ P_1$ production. Here, the symbol table attribute $\eta$ is passed between siblings from left to right, making this an inherited attribute. The synthesized symbol table from $S$ is passed as an inherited attribute to $P_1$, and then the final symbol table is passed up.
\begin{align*}
    P &\to S \ P_1 & & S.\eta = P.\eta \\
    &&& P_1.\eta = S.\eta_\mathit{res}\\
    & & & P.\eta_\mathit{res} = P_1.\eta_\mathit{res}\\
    P &\to S & & S.\eta = P.\eta \\
    & & & P.\eta_\mathit{res} = S.\eta_\mathit{res}
\end{align*}

\section{Overview}
\label{sec:overview}
We introduce Residual Attribute Grammars (RAGs)
and their semantic derivatives, lifting attribute evaluation to the incomplete parse trees produced during live LLM generation.
Our approach maps incomplete derivations to structured constraint formulas to ensure semantically invalid partial trees are disallowed.
In this section, we introduce the core mechanics of RAG and their semantic derivatives using a concrete expression language.

\subsection{Residual Relational Constraint System}
Traditional attribute grammars assume an evaluation context over fully concrete, static syntax trees. When an evaluator encounters an incomplete or \emph{partial} tree containing unexpanded regions, such as unknown future LLM generations, top-down and bottom-up attribute flows stall, as they are unable to reason about infinite completion spaces.

We lift the partial syntax tree into a many-sorted residual relational constraint system\footnote{We write these constraints using rewrite-rule notation, but they are interpreted declaratively.}. Syntactic structures and transfer functions are compiled into algebraic terms. Unexpanded structural non-terminals are represented as typed placeholder variables, or \emph{holes}. Instead of evaluating attributes functionally, we compile local semantic equations of the instantiated productions into a set of baseline term rules.

Consider the following tree with a non-terminal hole of sort $E$:
$$ t \Coloneqq \text{str } \textit{foo} = \Box_E $$
Our realizability goal is to determine if at least one concrete tree  $t_c$ derived from the incomplete tree $t$ exists such that $\llbracket t_c \rrbracket \downarrow \epsilon.\eta_\mathit{res} \neq \mathsf{Invalid}$ (i.e., that the root symbol table is not invalid). Given our incomplete expression $t$, the framework generates a tight residual relational constraint system representing the semantic properties of the given parse tree, which is then used to store and evaluate its attributes. To begin, the framework maps an attribute grammar's semantic rules to our residual system.
Fig.~\ref{fig:general_ag_compilation} provides the direct mapping from standard semantic rules to their corresponding many-sorted residual relational rewrite system representations.  Positions along a syntax tree traversal are tracked as integers $\pi \in \mathbb{N}^*$, with $\epsilon$ referring to the starting production.

\begin{figure}[t]
\centering
\small
\begin{tabular}{lll}
\hline
\textbf{Production $p \in P$} & \textbf{Semantic Equation $E_p$} & \textbf{ Relational Rule $\mathscr{P}_\pi\langle p \rangle $} \\ \hline
$\overset{0}{P} \to \overset{1}{S}$ & $S.\eta = P.\eta$ & $\pi.1.\eta \to \pi.\eta$ \\
          & $P.\eta_{\mathit{res}} = S.\eta_{\mathit{res}}$ & $\pi.\eta_{\mathit{res}} \to \pi.1.\eta_{\mathit{res}}$ \\ \hline
$\overset{0}{S} \to \overset{1}{T} \ \overset{2}{\mathit{id}} = \overset{3}{E}$ & $E.\eta = S.\eta$ & $\pi.3.\eta \to \pi.\eta$ \\
                           & $S.\eta_{\mathit{res}} = \mathsf{update}(S.\eta, \mathit{id}.\mathit{lex}, E.\tau, T.\tau_{\mathit{decl}})$ & $\pi.\eta_{\mathit{res}} \to \mathsf{update}(\pi.\eta, \pi.2.\mathit{lex}, \pi.3.\tau, \pi.1.\tau_{\mathit{decl}})$ \\ \hline
$\overset{0}{E} \to \overset{1}{E_1} + \overset{2}{E_2}$ & $E_1.\eta = E.\eta$ & $\pi.1.\eta \to \pi.\eta$ \\
                  & $E_2.\eta = E.\eta$ & $\pi.2.\eta \to \pi.\eta$ \\
                  & $E.\tau = \mathsf{promote}(E_1.\tau, E_2.\tau)$ & $\pi.\tau \to \mathsf{promote}(\pi.1.\tau, \pi.2.\tau)$ \\ \hline
\end{tabular}
\caption{Production mapping of several attribute grammar equations. An index of the production $i$ maps to local positional terms under a unified sorted signature.}
\label{fig:general_ag_compilation}
\end{figure}

\begin{figure*}[t]
\centering
\begin{subfigure}[t]{0.32\textwidth}
\centering
\resizebox{\linewidth}{!}{
\begin{tikzpicture}[
    >=Latex,
    level distance=1.5cm,
    sibling distance=1.6cm,
    edge from parent/.style={draw, thick},
    ast/.style={circle, draw, thick, minimum size=8mm, fill=gray!10, font=\Large},
    leaf/.style={rectangle, draw, thick, minimum size=8mm, rounded corners, fill=gray!10, font=\ttfamily\large, inner sep=6pt},
    partial/.style={rectangle, draw=orange!90!black, thick, dashed, minimum size=8mm, rounded corners, fill=orange!10, font=\Large, text=orange!90!black},
    label distance=-0.2cm,
    every label/.style={minimum size=10pt, text=white, fill=black, font=\small, in front of path, inner sep=0pt},
]

\node[ast, label={above right:$\epsilon$}] {$P$}
    child {node[ast, label={above right:1}] {$S$}
        child {node[ast, label={above right:1.1}] {$T$}
            child {node[leaf, label={above right:1.1.1}] {str}}
        }
        child {node[leaf, label={above right:1.2}] {foo}}
        child {node[leaf] {=}}
        child {node[partial, label={above right:1.3}] {$\Box_E$} edge from parent[dashed, thick, draw=orange!90!black]}
    };
\end{tikzpicture}}
\caption{$\Box_E$ : unexpanded hole of sort $E$}
\label{sfig:partial_parse_tree}
\end{subfigure}
\hfill
\begin{subfigure}[t]{0.32\textwidth}
\centering
\resizebox{\linewidth}{!}{
\begin{tikzpicture}[
    >=Latex,
    level distance=1.5cm,
    sibling distance=1.6cm,
    edge from parent/.style={draw, thick},
    ast/.style={circle, draw, thick, minimum size=8mm, fill=gray!10, font=\Large},
    leaf/.style={rectangle, draw, thick, minimum size=8mm, rounded corners, fill=gray!10, font=\ttfamily\large, inner sep=6pt},
    ast_new/.style={circle, draw, thick, minimum size=8mm, fill=orange!20, font=\Large},
    leaf_new/.style={rectangle, draw, thick, minimum size=8mm, rounded corners, fill=orange!20, font=\ttfamily\large, inner sep=6pt},
    label distance=-0.2cm,
    every label/.style={minimum size=10pt, text=white, fill=black, font=\small, in front of path, inner sep=0pt},
]

\node[ast, label={above right:$\epsilon$}] {$P$}
    child {node[ast, label={above right:1}] {$S$}
        child {node[ast, label={above right:1.1}] {$T$}
            child {node[leaf, label={above right:1.1.1}] {str}}
        }
        child {node[leaf, label={above right:1.2}] {foo}}
        child {node[leaf] {=}}
        child {node[ast_new, label={above right:1.3}] {$E$}
            child {node[leaf_new, label={above right:1.3.1}] {"bar"}}
        }
    };
\end{tikzpicture}}
\caption{$E \to \mathsf{string\_const}$ : $E.\tau=\mathsf{String}$ (valid)}
\label{sfig:valid_branch}
\end{subfigure}
\hfill
\begin{subfigure}[t]{0.32\textwidth}
\centering
\resizebox{\linewidth}{!}{
\begin{tikzpicture}[
    >=Latex,
    level distance=1.5cm,
    sibling distance=1.6cm,
    edge from parent/.style={draw, thick},
    ast/.style={circle, draw, thick, minimum size=8mm, fill=gray!10, font=\Large},
    leaf/.style={rectangle, draw, thick, minimum size=8mm, rounded corners, fill=gray!10, font=\ttfamily\large, inner sep=6pt},
    ast_new/.style={circle, draw, thick, minimum size=8mm, fill=orange!20, font=\Large},
    leaf_new/.style={rectangle, draw, thick, minimum size=8mm, rounded corners, fill=orange!20, font=\ttfamily\large, inner sep=6pt},
    partial/.style={rectangle, draw=orange!90!black, thick, dashed, minimum size=8mm, rounded corners, fill=orange!10, font=\Large, text=orange!90!black},
    label distance=-0.2cm,
    every label/.style={minimum size=10pt, text=white, fill=black, font=\small, in front of path, inner sep=0pt},
]

\node[ast, label={above right:$\epsilon$}] {$P$}
    child {node[ast, label={above right:1}] {$S$}
        child {node[ast, label={above right:1.1}] {$T$}
            child {node[leaf, label={above right:1.1.1}] {str}}
        }
        child {node[leaf, label={above right:1.2}] {foo}}
        child {node[leaf] {=}}
        child {node[ast_new, label={above right:1.3}] {$E$}
                child {node[ast_new, label={above right:1.3.1}] {$E$}
                    child {node[leaf_new, label={above right:1.3.1.1}] {"bar"}}
                }
                child {node[leaf_new] {+}}
                child {node[partial, label={above right:1.3.2}] {$\Box_E$} edge from parent[dashed, thick, draw=orange!90!black]}
        }
    };
\end{tikzpicture}}
\caption{$E \to E_1 + E_2$ : $E.\tau=\mathsf{promote}(\mathsf{String},\cdot)=\bot$ (pruned)}
\label{sfig:pruned_branch}
\end{subfigure}
\caption{Partial parse trees for \texttt{str foo = [e]}: (a) the unexpanded structural hole $\Box_E$ of sort $E$; (b, c) its two production choices after the derivative consumes the lexeme \texttt{"bar"}, selected by the structural choice variable $\Omega_{1.3}$.}
\label{fig:parse_trees}
\label{fig:refinement_branches}%
\end{figure*}
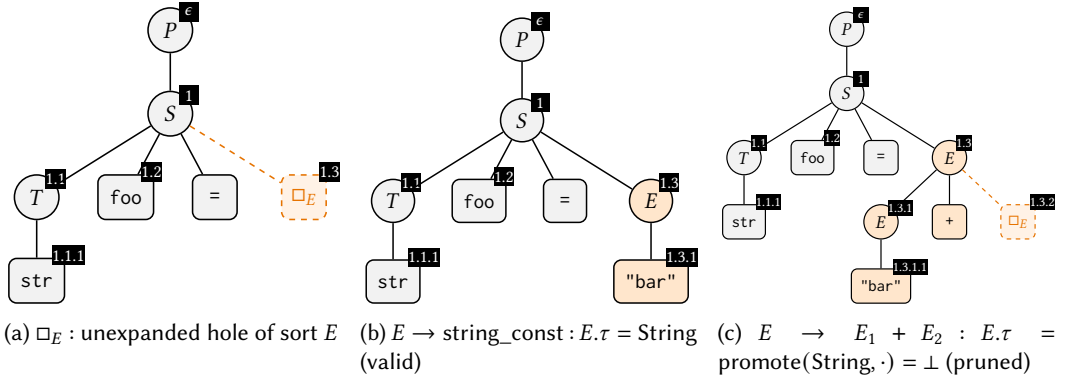

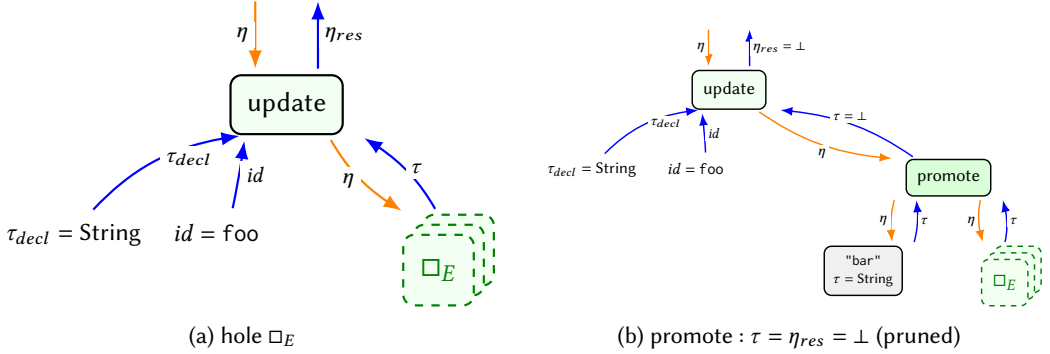
\begin{figure*}[t]
\centering
\begin{subfigure}[t]{0.48\textwidth}
\centering
\resizebox{\linewidth}{!}{
\begin{tikzpicture}[
    >=Latex,
    node distance=2.5cm and 3cm,
    ast_box/.style={rectangle, draw, thick, rounded corners, fill=green!5, font=\sffamily, inner sep=6pt, minimum width=0.5cm, align=center},
    ast_circle/.style={rectangle, draw=green!50!black, thick, rounded corners, dashed, minimum size=9mm, fill=green!5, font=\Large\bfseries\itshape, text=green!50!black},
    attr/.style={text=black, font=\small\sffamily, align=center},
    down_flow/.style={->, thick, orange, shorten >=2pt, shorten <=2pt},
    up_flow/.style={->, thick, blue, shorten >=2pt, shorten <=2pt},
    edge_label/.style={fill=white, inner sep=2pt, font=\footnotesize\sffamily, text=black}
]

\node[ast_box] (update) {$\mathsf{update}$};

\node[above=1.0cm of update] (top_anchor) {};

\node[ast_circle, below right=1.0cm and 1.0cm of update] (e) {$\Box_E$};
\node[ast_circle, below left=-0.8cm and -0.8cm of e] (e1) {$\Box_E$};
\node[ast_circle, below left=-0.8cm and -0.8cm of e1] (e2) {$\Box_E$};

\node[attr, below left=1.0cm and 1.0cm of update] (decl1) {$\tau_{decl}=\mathsf{String}$};

\node[attr, below left=1.0cm and -0.5cm of update] (decl2) {$\mathit{id}=\texttt{foo}$};

\draw[down_flow] ([xshift=-0.4cm]top_anchor.south) -- ([xshift=-0.4cm]update.north) node[midway, left, edge_label] {$\eta$};
\draw[up_flow] ([xshift=0.4cm]update.north) -- ([xshift=0.4cm]top_anchor.south) node[midway, right, edge_label] {$\eta_{res}$};

\draw[up_flow] ([xshift=0.2cm]decl1.north) to[bend left=20] node[midway, right, edge_label] {$\tau_{decl}$} ([xshift=0.2cm]update.south west);
\draw[up_flow] ([xshift=0.2cm]decl2.north) to[bend right=2] node[midway, right, edge_label] {$\mathit{id}$} ([xshift=0.2cm]update.south west);

\draw[down_flow] ([xshift=-0.2cm]update.south east) to[bend right=15] node[midway, left, edge_label] {$\eta$} ([xshift=-0.2cm]e.north west);
\draw[up_flow] ([xshift=0.2cm]e.north west) to[bend right=15] node[midway, right, edge_label] {$\tau$} ([xshift=0.2cm]update.south east);
\end{tikzpicture}}
\caption{hole $\Box_E$}
\label{fig:relational_summary}
\end{subfigure}
\hfill
\begin{subfigure}[t]{0.48\textwidth}
\centering
\resizebox{\linewidth}{!}{
\begin{tikzpicture}[
    >=Latex,
    node distance=2.5cm and 3cm,
    ast_box/.style={rectangle, draw, thick, rounded corners, fill=green!5, font=\sffamily, inner sep=6pt, minimum width=0.5cm, align=center},
    ast_box_new/.style={rectangle, draw, thick, rounded corners, fill=green!15, font=\sffamily, inner sep=6pt, minimum width=0.5cm, align=center},
    leaf_frozen/.style={rectangle, draw, thick, rounded corners, fill=gray!12, font=\footnotesize\sffamily, inner sep=5pt, align=center},
    ast_circle/.style={rectangle, draw=green!50!black, thick, rounded corners, dashed, minimum size=9mm, fill=green!5, font=\Large\bfseries\itshape, text=green!50!black},
    attr/.style={text=black, font=\small\sffamily, align=center},
    down_flow/.style={->, thick, orange, shorten >=2pt, shorten <=2pt},
    up_flow/.style={->, thick, blue, shorten >=2pt, shorten <=2pt},
    edge_label/.style={fill=white, inner sep=1.5pt, font=\footnotesize\sffamily, text=black}
]

\node[ast_box] (update) {$\mathsf{update}$};
\node[above=0.85cm of update] (top_anchor) {};

\node[ast_box_new, below right=0.95cm and 2.7cm of update] (promote1) {$\mathsf{promote}$};

\node[leaf_frozen, below left=0.95cm and 0.05cm of promote1] (barleaf) {\texttt{"bar"}\\$\tau=\mathsf{String}$};

\node[ast_circle, below right=0.95cm and 0.05cm of promote1] (e) {$\Box_E$};
\node[ast_circle, below left=-0.8cm and -0.8cm of e] (e1) {$\Box_E$};
\node[ast_circle, below left=-0.8cm and -0.8cm of e1] (e2) {$\Box_E$};

\node[attr, below left=0.9cm and 0.95cm of update] (decl1) {$\tau_{decl}=\mathsf{String}$};
\node[attr, below left=0.9cm and -0.75cm of update] (decl2) {$\mathit{id}=\texttt{foo}$};

\draw[down_flow] ([xshift=-0.4cm]top_anchor.south) -- ([xshift=-0.4cm]update.north) node[midway, left, edge_label] {$\eta$};
\draw[up_flow] ([xshift=0.4cm]update.north) -- ([xshift=0.4cm]top_anchor.south) node[midway, right, edge_label] {$\eta_{res}=\bot$};

\draw[up_flow] ([xshift=0.2cm]decl1.north) to[bend left=20] node[midway, right, edge_label] {$\tau_{decl}$} ([xshift=0.2cm]update.south west);
\draw[up_flow] ([xshift=0.2cm]decl2.north) to[bend right=2] node[midway, right, edge_label] {$\mathit{id}$} ([xshift=0.2cm]update.south west);

\draw[down_flow] ([xshift=-0.2cm]update.south east) to[bend right=15] node[midway, below, edge_label] {$\eta$} ([xshift=-0.2cm]promote1.north west);
\draw[up_flow] ([xshift=0.2cm]promote1.north west) to[bend right=15] node[midway, above, edge_label] {$\tau=\bot$} ([xshift=0.2cm]update.south east);

\draw[down_flow] ([xshift=-0.2cm]promote1.south west) to[bend right=12] node[midway, left, edge_label] {$\eta$} ([xshift=-0.2cm]barleaf.north east);
\draw[up_flow] ([xshift=0.2cm]barleaf.north east) to[bend right=12] node[midway, right, edge_label] {$\tau$} ([xshift=0.2cm]promote1.south west);

\draw[down_flow] ([xshift=-0.2cm]promote1.south east) to[bend right=15] node[midway, left, edge_label] {$\eta$} ([xshift=-0.2cm]e.north west);
\draw[up_flow] ([xshift=0.2cm]e.north west) to[bend right=15] node[midway, right, edge_label] {$\tau$} ([xshift=0.2cm]promote1.south east);
\end{tikzpicture}}
\caption{$\mathsf{promote}$ : $\tau=\eta_{res}=\bot$ (pruned)}
\label{fig:relational_summary_addition}
\end{subfigure}
\caption{Semantic attribute evaluation over \Cref{fig:parse_trees}: inherited attributes ($\eta$) flowing downward (orange), synthesized attributes ($\tau, \tau_{decl}, \mathit{id}, \eta_{res}$) upward (blue).}
\label{fig:relational_summaries}
\end{figure*}

Traversing $t$ and applying the relational rules per production $\mathscr{P}_\pi\langle p \rangle$ reveals a compiled residual relational constraint system $\mathscr{C}$ at the index $\epsilon$ for the top-level production $P$ given the prefix \texttt{str foo =}\footnote{The relational system is simplified for this example.}:
\begin{align*}
\mathscr{C}_\epsilon\langle\Box_{P \backslash \texttt{str foo =}}\rangle = \begin{cases}
    \epsilon.\eta_{\mathit{res}} &\to \mathsf{update}(\{\}, \text{``foo''}, {\color{orange!90!black}1.3.\tau}, \mathsf{String}) \\
    {\color{orange!90!black}1.3}.\eta &\to 1.\eta \\
\end{cases} \cup \mathscr{C}_{1.3}\langle\Box_{E}\rangle
\end{align*}
At position $1.3$, the tree recurses into the $E$ production, where the unexpanded hole is captured by a placeholder $1.3.\tau$ (Fig.~\ref{sfig:partial_parse_tree}). The recursive $E$ production's system is denoted as $\mathscr{C}_{1.3}\langle\Box_E\rangle$.  Attributes whose evaluation paths depend on an unexpanded structural hole are deferred to the residual system. Figure~\ref{fig:relational_summary} illustrates this: the inherited symbol table $\eta$ is threaded downward into the $\mathsf{update}$ node, while the synthesized $\tau$ and $\eta_{res}$, both dependent on the hole $\Box_E$, remain symbolic. Given a complete attribute tree, the relational constraint system could be simplified to contain only a final $\epsilon.\eta_{res}$.

The syntax tree $t$ with structural holes represents an infinite number of concrete trees as $\Box_E$ may expand into \emph{any} production of $E$. To align with the paradigm of constrained decoding, which restricts a decoder to continuations admitting at least one valid completion, we evaluate these structural holes by asking if \emph{there exists at least one concrete tree} such that the top-level predicate is met. Consequently, the problem of identifying a valid concrete tree reduces entirely to discovering a single satisfiable computation path within the relational constraint system.

A non-terminal grammar symbol, such as $E$, can realize to one of many production rules. To distinguish between productions on the same non-terminal symbol, our frameworks assigns a set of guarded relational rules whose non-determinism is carried by a structural choice variable $\Omega$.
In the case of $E$ and the corresponding hole $\Box_E$, there are 5 distinct choices, each guarded by $\Omega_\pi$ via the guard operator $[\,\Omega_\pi = i\,] \triangleright \mathit{Rule}$:

{\small $$
\mathscr{C}_\pi\langle\Box_E\rangle=(\Omega_\pi \in \{1,\dots,5\}) \;\cup\;
\begin{cases}
[\Omega_\pi = 1] \triangleright\ \pi.\Box_E.\tau \to \mathsf{String}\\
[\Omega_\pi = 2] \triangleright\ \pi.\Box_E.\tau \to \mathsf{Int} \\
[\Omega_\pi = 3] \triangleright\ \pi.\Box_E.\tau \to \mathsf{Float} \\
[\Omega_\pi = 4] \triangleright\ \pi.\Box_E.\tau \to \mathsf{lookup}(\pi.\eta, \pi.\mathit{lex}) \\
[\Omega_\pi = 5] \triangleright
\left\{ \begin{aligned}
\pi.1.\Box_E.\eta &\to \pi.\eta \\
\pi.2.\Box_E.\eta &\to \pi.\eta \\
\pi.\Box_E.\tau &\to \mathsf{promote}(\pi.1.\Box_E.\tau, \pi.2.\Box_E.\tau)
\end{aligned} \right.
\end{cases}
$$}

\noindent
In these relational rules, the holes are indexed by a placeholder $\pi$ prefix representing a relational position instead of an absolute position prefix. The relational indices allow us to (1) reuse hole definitions regardless of where they exist on a tree, and (2) represent recursive rules.

\paragraph{Solving with oracles} Solving for these relational rules is done through an abstract oracle $\vdash_\textit{oracle}$ (e.g., SMT solver). The oracle receives the relational constraint system, and evaluates the predicate. If a \emph{valid path} exists such that the predicate is met, the relational constraint system, and the tree and prefix it was compiled from, are valid.
\textsc{AGDecode} is sound and complete relative to an ideal oracle, however, because solvers may return unknown on complex inductive queries, the implementation must decide how to handle uncertainty.
With practical solvers, conservative handling of unknown yields sound but incomplete pruning; our prototype also supports a speculative mode that admits timed-out tokens and later revalidates longer prefixes, trading immediate soundness for progress.

\subsection{Symbolic Refinement and Rule Transformations}

As generation progresses, structural holes are expanded into explicit grammar productions through a state transition relation termed \emph{refinement}. Refinement is used to systematically narrow holes to represent smaller, more specific tree spaces and continuously transform the residual relational constraint system to represent their semantics.
Standard CFG derivatives, extended from Brzozowski derivatives \cite{brzozowski_derivatives_1964}, can be applied to yield the set of valid string suffixes given a prefix \cite{might_parsing_2011}. We lift this process to partial syntax trees.

The process of consuming characters is mapped to the process of \emph{refining} holes. Suppose we have the tree $t$ and the lexeme \texttt{"bar"} is generated.
In this case, the system determines that the structural hole $\Box_E$ is refined into $\Box_{E \backslash \texttt{"bar"}}$, which is any production of $E$ where the leftmost characters are \texttt{"bar"}. While some productions of $E$ are constant, such as $E \to \text{string\_const}$  (Fig.~\ref{sfig:valid_branch}),
the the addition production $E \to E_1 + E_2$ (Fig.~\ref{sfig:pruned_branch}) could recurse infinitely through $E_1$.
Standard CFG derivative processes rely on thunk-ing and fixed-point algorithms to represent and reason about these recursive properties. \cite{might_parsing_2011}. In \textsc{AGDecode}, infinite recursion is represented by structural hole placeholders and evaluated through an  oracle.
During refinement, the relational constraint system undergoes an explicit rule transformation. These rules match the required structural rules for the given replacement as governed by the attribute grammar. Fig.~\ref{fig:rule_transformation} demonstrates this expansion.
\begin{figure}[h]
\centering
{\small$$\mathscr{C}_\epsilon\langle\Box_{P \backslash \texttt{str foo = "bar"}}\rangle = \begin{cases}
    \epsilon.\eta_{\mathit{res}} &\to \mathsf{update}(\{\}, \text{``foo''}, {\color{orange!90!black}1.3.\tau}, \mathsf{String}) \\
    {\color{orange!90!black}1.3}.\eta &\to 1.\eta \\
\end{cases} \cup \mathscr{C}_{1.3}\langle\Box_{E\backslash\texttt{"bar"}} \rangle$$}
{\small $$\mathscr{C}_\pi\langle\Box_{E\backslash\texttt{"bar"}}\rangle=
\begin{cases}
    [\Omega_\pi = 1] \triangleright\ \pi.\Box_{E\backslash\texttt{"bar"}}.\tau \to \mathsf{String}\\
    [\Omega_\pi = 2] \triangleright\ \pi.\Box_{E\backslash\texttt{"bar"}}.\tau \to \mathsf{Int} \\
    [\Omega_\pi = 3] \triangleright\ \pi.\Box_{E\backslash\texttt{"bar"}}.\tau \to \mathsf{Float} \\
    [\Omega_\pi = 4] \triangleright\ \pi.\Box_{E\backslash\texttt{"bar"}}.\tau \to \mathsf{lookup}(\pi.\eta, \pi.\mathit{lex}) \\
    [\Omega_\pi = 5] \triangleright \begin{aligned}[t]
        &\left\{ \begin{aligned}
            \pi.1.\Box_{E\backslash\texttt{"bar"}}.\eta &\to \pi.\eta \\
            \pi.2.\Box_E.\eta &\to \pi.\eta \\
            \pi.\Box_{E\backslash\texttt{"bar"}}.\tau &\to \mathsf{promote}(\pi.1.\Box_{E\backslash\texttt{"bar"}}.\tau, \pi.2.\Box_E.\tau)
        \end{aligned} \right. \\
        &\quad \cup \mathscr{C}_{\pi.1}\langle\Box_{E\backslash\texttt{"bar"}}\rangle \cup \mathscr{C}_{\pi.2}\langle\Box_{E}\rangle
    \end{aligned}
\end{cases}$$}
\caption{Transformation of the residual relational constraint system during a structural refinement step at position $\pi = 1.3$. The only valid semantic production is $\Omega_\pi=1$.}
\label{fig:rule_transformation}
\end{figure}
The evaluation oracle then processes $\mathscr{C}_\epsilon\langle\Box_{P\backslash \texttt{str foo = "bar"}\rangle}$. The structural expansion introduces a deterministic type mismatch; the result of $\mathsf{promote}$ will never be $\mathsf{String}$, as our semantics disallow string addition. Thus, the top-level query, which expects an expression of the type $\mathsf{String}$, is unsatisfiable by the oracle, as is the case with the productions for the integer, floating-point, and lookup productions ($\Omega_\pi = 2\dots5$)
demonstrated in Fig.~\ref{fig:relational_summary_addition}. However, the string literal production ($\Omega_\pi = 1$) is semantically valid, allowing the new prefix to be accepted. The semantic validity problem is both existentially and universally quantified, in that we want to prove that there exists at least one valid semantic path (i.e., selections of $\Omega_\pi$) such that no semantic errors can occur.

\subsubsection{Partial Lexemes}
\label{sec:derivative_overview}
Evaluating semantic rules over an incomplete lexeme (e.g., a half-typed variable identifier) introduces an infinite search space, as the prefix could theoretically resolve to an infinite number of distinct lexemes.
To ensure decidability and monotonic convergence, our framework introduces a strict division between partial lexemes and concrete realized lexemes. We replace unknown lexical synthesized attributes with two distinct structural forms: (1)
 \textbf{Lexical Holes ($\Box_{c\backslash w}$):} An over-approximation of the constraint space for an incomplete prefix string $w$, or (2) \textbf{Solid Terminal Nodes ($\blacksquare_{c\backslash w}$):} The exact, localized constraint compiled once a token boundary is reached and the lexeme is frozen.
The grammar enforces a  soundness constraint requiring the abstract prefix to logically subsume all possible future concrete completions, expressed as the refinement chain $\Box_{c\backslash \epsilon} \sqsubseteq \Box_{c\backslash w} \sqsubseteq \blacksquare_{c\backslash w}$.

\subsection{Generalizing Semantics as Constrained Horn Clauses}

For simple, finite constraint domains, it is entirely feasible to implement bespoke fixed-point solvers or custom abstract interpreters to resolve the relational summaries. However, our design gives us the ability to lift summaries from domain-specific transfers of information into general first- and higher-order logic.
We achieve this by lowering the recursive relational summaries of our attribute grammar into CHCs.

A CHC is a fragment of first-order logic \cite{beklemishev_horn_2015} taking the form of an implication:
\begin{align*}
    \forall \vec{v} . \left( \phi \land P_1(\vec{x}_1) \land \dots \land P_n(\vec{x}_n) \implies H(\vec{x}) \right),
\end{align*}
where $\phi$ is a formula in a background theory (e.g., linear integer arithmetic, uninterpreted functions), $P_i$ are uninterpreted relational predicates representing our sub-trees, and $H$ is the head predicate.

In this CHC-backed formulation, the abstract transfers are expressed as relations.
We define an uninterpreted predicate for each non-terminal.
For instance, our earlier sum promotion rule translates naturally into a Horn clause, joining the evaluation of its children with the logic of the $\mathsf{promote}$ relation:
\begin{align*}
    \forall \eta, \tau_1, \tau_2, \tau_{res} . \Big( & \textsf{SumNode}_{left}(\eta, \tau_1) \land \textsf{SumNode}_{right}(\eta, \tau_2) \\
    & \land \mathsf{promote}(\tau_1, \tau_2) = \tau_{res} \\
    & \implies \textsf{SumNode}_{parent}(\eta, \tau_{res}) \Big)
\end{align*}

By utilizing SMT background theories like the Theory of Arrays, Bitvectors, or Datatypes, this architecture can be  extended to enforce highly complex semantic properties on the fly.

\section{Declarative Semantics of Incomplete Residual Attribute Grammars}
\label{sec:semantic_ags}

In this section, we formalize a declarative framework for reasoning about semantic attribute constraints over incomplete syntax structures.
We begin by augmenting classical attribute grammars with a typed term algebra. (Sec.~\ref{sec:ag_transfer_functions}).
We introduce partial syntax trees to account for incomplete trees and embed their corresponding semantic configurations into static systems of residual relational constraints (Sec.~\ref{sec:incomplete_trees}).
We provide a global compilation function mapping partial tree geometry directly to these constraint equations (Sec.~\ref{sec:narrowing}).
Finally, we prove that evaluating satisfiability over these partial constraint spaces is strictly sound and monotonic with respect to traditional evaluation over complete ground trees (Sec.~\ref{sec:ag_theorems}).
Crucially, this declarative formulation replaces dynamic operational tree-refining with static logical judgments, allowing an automated reasoning oracle to prune semantically invalid search spaces early.

\subsection{Residual Attribute Grammars}
\label{sec:ag_transfer_functions}

Formally, a CFG is a 4-tuple $G = (N, \Sigma, P, S)$, where $N$ is the finite set of non-terminals, $\Sigma$ is the set of terminals disjoint from $N$, $P \subseteq N \times (N \cup \Sigma)^*$ is the set of productions, and $S \in N$ is the start symbol. Let $V = N \cup \Sigma$ denote the complete vocabulary. We assume a lower-level physical alphabet $\Sigma_{\mathit{char}}$ (e.g., ASCII characters) and a lexical mapping function $\mathcal{L} : \Sigma \to \mathcal{P}(\Sigma_{\mathit{char}}^*)$ associating each terminal with a regular language of characters.

An AG~\cite{knuth_semantics_1968, aho_compilers_2006} is a 4-tuple $\mathit{AG} = (G, A, \mathcal{D}, E)$ where $G$ is the underlying CFG, $A$ associates each symbol $X \in V$ with a finite set of attributes $A(X)$, $\mathcal{D}$ maps each attribute $a \in A(X)$ to a domain $D(a)$, and $E$ associates each production $p \in P$ with a finite set of semantic equations $E_p$. The set $A(X)$ partitions into disjoint subsets: synthesized attributes $\mathbf{S}(X)$ and inherited attributes $\mathbf{I}(X)$. By convention, $\mathbf{I}(S) = \emptyset$, and $\forall c \in \Sigma.\ \mathbf{I}(c) = \emptyset$.

A production $p \in P$ takes the form $X_0 \to X_1 X_2 \dots X_n$. The set $E_p$ defines the computation of specific attributes within the local context of $p$.

To reason about semantic constraints symbolically, we augment standard AGs with a typed term algebra in place of opaque computable functions, creating a Residual Attribute Grammar (RAG).

\begin{definition}[Residual Attribute Grammar]
A \emph{RAG} is a 6-tuple $\mathit{RAG} = (G, A, \mathcal{D}, \mathcal{F}, \Sigma_{\mathit{RAG}}, E)$, where:
\begin{itemize}
    \item $G, A, \mathcal{D}$ mirror the standard AG definitions.
    \item $\mathcal{F}$ is a many-sorted signature of computation symbols. Each function $f \in \mathcal{F}$ possesses a strictly defined type $f : d_1 \times \dots \times d_m \to d_0$, where $d_k \in \mathcal{D}$.
    \item $\Sigma_{\mathit{RAG}}$ is the expanded algebraic signature including $\mathcal{F}$ alongside unary attribute projection symbols. For every $X \in V$ and $a \in A(X)$, $\Sigma_{\mathit{RAG}}$ contains a symbol $a : X \to D(a)$.
    \item $E$ maps each production $p \in P$ to a finite set of attribute evaluation rules $E_p$.
\end{itemize}
\end{definition}

Let $\mathcal{V}_p = \{x_0, x_1, \dots, x_n\}$ be a set of typed variables where each $x_k$ has sort $X_k$. The set $E_p$ computes attributes using the many-sorted term algebra $T_{\Sigma_{\mathit{RAG}}}(\mathcal{V}_p)$. Each evaluation rule in $E_p$ takes the syntactic form $x_i.a = t_{\mathit{rhs}}$, where:
\begin{enumerate}[label=(\alph*)]
    \item the target term $x_i.a$ is uniquely defined by exactly one rule in $E_p$,
    \item the target attribute $a$ is either a synthesized attribute of the left-hand side ($i = 0 \land a \in \mathbf{S}(X_0)$) or an inherited attribute of a right-hand side symbol ($1 \le i \le n \land a \in \mathbf{I}(X_i)$),
    \item $t_{\mathit{rhs}} \in T_{\Sigma_{\mathit{RAG}}}(\mathcal{V}_p)_{D(a)}$ is a well-typed term evaluating to the target domain.
\end{enumerate}

The set of all \emph{complete} syntax trees in $\mathit{RAG}$ is the set of well-defined terms in the term algebra $t \in T_{\Sigma_{\mathit{RAG}}}(\emptyset)$.
We address tree nodes uniquely via positions $\pi \in \mathbb{N}^*$. The root is denoted by the empty sequence $\epsilon$, and $\pi.i$ denotes the $i$-th child of the node at $\pi$. Let $t|_\pi$ denote the subterm of $t$ at position $\pi$. Attributes on a complete tree are evaluated in a traditional manner.

\begin{definition}[Semantic Evaluation of a Complete Tree]
The evaluation of an attribute $a$ at position $\pi$ on a \emph{complete} syntax tree $t \in T_{\Sigma_{\mathit{RAG}}}(\emptyset)$, defined as the operation
$\llbracket t \rrbracket \downarrow \pi.a$
is sound and complete. The codomain of evaluation is strictly $D(a)$.
\end{definition}
The definition notably encapsulates some functional definition that correspond to each signature $f \in \mathcal{F}$ and assumes all are soundly and completely evaluated. This axiom is set up to later ensure that abstract evaluation over \emph{incomplete} trees are sound and complete. We define the \emph{semantic validity} of a tree as the evaluated result of some synthesized boolean attribute $\epsilon.\mathit{valid}$.

\subsection{Partial Syntax Trees and Residual Constraint Systems}
\label{sec:incomplete_trees}
To represent unexpanded productions and unrealized lexemes, we introduce a family of vocabulary holes $\mathcal{H}$ as placeholder symbols. For any symbol $X$ in our complete vocabulary $V = N \cup \Sigma$, the term $\Box_X$ represents an uninstantiated non-terminal or an unexpanded terminal hole of sort $X$. These are used in both syntactic and semantic representations of a tree.
\subsubsection{Syntax: Partial Derivation Trees}

\begin{definition}[Partial Derivation Tree]
A \emph{Partial Derivation Tree} is a well-typed term in the extended term algebra $T_{\Sigma_{\mathit{RAG}}}(\mathcal{H})$. A tree is \emph{complete} if it contains no holes (i.e., it is a ground term $t \in T_{\Sigma_{\mathit{RAG}}}(\emptyset)$).
Substitution $t[\Box_X \leftarrow t']_\pi$ denotes the replacement of the subtree at position $\pi$ (which must be $\Box_X$) with the term $t'$.
\end{definition}

To formally ground structural refinement, we define the homomorphic extension $\hat{\sigma} : T_{\Sigma_{\mathit{RAG}}}(\mathcal{H}) \to T_{\Sigma_{\mathit{RAG}}}(\mathcal{H})$ of a base substitution $\sigma : \mathcal{H} \to T_{\Sigma_{\mathit{RAG}}}(\mathcal{H})$ by structural induction:
\begin{align*}
    \hat{\sigma}(\Box{}_X) &= \sigma(\Box{}_X) \quad \text{for } \Box{}_X \in \mathcal{H} \\
    \hat{\sigma}(c) &= c \quad \text{for nullary } c \in \Sigma_{\mathit{RAG}} \\
    \hat{\sigma}(f(t_1, \dots, t_n)) &= f(\hat{\sigma}(t_1), \dots, \hat{\sigma}(t_n)) \quad \text{for } f \in \Sigma_{\mathit{RAG}}
\end{align*}

\begin{definition}[Refinement and Concretization of Partial Derivation Tree]
Let $t_1, t_2 \in T_{\Sigma_{\mathit{RAG}}}(\mathcal{H})$. Tree $t_2$ \emph{refines} $t_1$, denoted $t_1 \sqsubseteq t_2$, iff there exists a well-typed substitution $\sigma$ such that $\hat{\sigma}(t_1) = t_2$.
The \emph{concretization function} $\gamma$ maps a partial derivation tree $t$ to the infinite set of complete, ground trees that refine it:
$$    \gamma(t) = \{ t' \in T_{\Sigma_{\mathit{RAG}}}(\emptyset) \mid t \sqsubseteq t' \}$$
\end{definition}

\subsubsection{Semantics: Residual Relational Constraint Systems}
To represent the accompanying semantics of incomplete trees,
we lift local semantic equations into a unified \emph{Residual Relational Constraint System }.

\begin{definition}[Residual Relational Constraint System ]

A \emph{Residual Relational Constraint System } $\mathcal{R}$ is a finite set of sorted term relational constraint rules tracking open relational constraints over attribute paths. Let $A = \bigcup_{X \in V} A(X)$ be the set of all semantic attributes. We define an \emph{Attribute Target} as a projection coordinate pair $\pi.a$, where $\pi \in \mathbb{N}^*$ and $a \in A$ is a semantic attribute of $X$. Each attribute $a$ has a designated semantic sort/domain $D(a) \in \mathcal{D}$.
Formally, a well-sorted residual relational constraint system  $\mathcal{R}$ is a relation: $\mathcal{R} : (\mathbb{N}^* \times A) \to \mathcal{P}(T_{\Sigma_{\mathit{RAG}}})$
where all rules $l \to r \in \mathcal{R}$ satisfy the invariants that:
\begin{enumerate}
    \item the left-hand side is a bare attribute projection path: $l = \pi.a$ where $\pi \in \mathbb{N}^*$ and $a \in A$,
    \item the right-hand side term must ground to the same semantic sort as its target attribute domain: $\text{sort}(r) = D(a)$.
\end{enumerate}
\end{definition}

The relational constraint system isolates the semantic constraints of unexpanded holes, deferring their evaluation until structural concretization. It acts as a relational semantic interface bounded to the partial syntax tree.
The holes in the relational constraint systems represent uninterpreted relational interfaces to all concrete evaluations over a traditional attribute grammar.
Just with the syntactical trees, the ruleset can be extended to those without holes through refinement and concretization.

We define the homomorphic rule extension $\hat{\sigma}_{\mathcal{R}} : \mathcal{R} \to \mathcal{R}$ of a well-sorted rule substitution $\sigma_{\mathcal{R}} : \mathcal{H} \to T_{\Sigma_{\mathit{RAG}}}(\mathcal{H})$ by structural induction over the right-hand side terms of the system. Let $l = \pi.a$ be an attribute target:
$$    \hat{\sigma}_{\mathcal{R}}(l \to \Box_X.a) = (l \to \sigma_{\mathcal{R}}(\hat{\sigma}_{\mathcal{R}}(\Box_X).a))$$
$$    \hat{\sigma}_{\mathcal{R}}(l \to c) = (l \to c) \quad \text{for nullary domain constants } c \in \Sigma_{\mathit{RAG}}$$
$$    \hat{\sigma}_{\mathcal{R}}(l \to f(t_1, \dots, t_n)) = (l \to f(\hat{\sigma}_{\mathcal{R}}(t_1), \dots, \hat{\sigma}_{\mathcal{R}}(t_n))) \quad \text{for functions } f \in \mathcal{F}$$

\begin{definition}[Refinement and Concretization of Residual Relational Constraint Trees]
Let $\mathcal{R}_1, \mathcal{R}_2$ be two residual relational constraint systems. $\mathcal{R}_2$ semantically refines $\mathcal{R}_1$, denoted $\mathcal{R}_1 \sqsubseteq_{\mathcal{R}} \mathcal{R}_2$, if and only if there exists a well-sorted rule substitution $\sigma_{\mathcal{R}}$ such that the normalized homomorphic transformation matches exactly:
$$   \mathcal{R}_1 \sqsubseteq_{\mathcal{R}} \mathcal{R}_2 \iff \exists \sigma_{\mathcal{R}}.\;\mathcal{R}_2 = \hat{\sigma}_{\mathcal{R}}(\mathcal{R}_1)$$
A residual relational constraint system $\mathcal{R}_c$ is complete or ground if and only if it contains no relational placeholder hole terms (i.e., its active set of unresolved placeholders is empty, $\mathcal{H}(\mathcal{R}_c) = \emptyset$)
The rule concretization function $\gamma_{\mathcal{R}}$ maps a partial relational constraint system $\mathcal{R}$ to the bounded set of all complete, ground relational constraint systems that refine it:
$$    \gamma_{\mathcal{R}}(\mathcal{R}) = \{ \mathcal{R}_c \mid \mathcal{R} \sqsubseteq_{\mathcal{R}} \mathcal{R}_c \;\land\; \mathcal{H}(\mathcal{R}_c) = \emptyset \}$$
\end{definition}

A relational constraint system may contain recursive and non-deterministic rules. Thus, we
defer the proof of existence to an abstract evaluation oracle $\vdash_{oracle}$. In implementation, this bounds the required behavior without mandating a specific fixed-point mechanism.

\begin{axiom}[Abstract Oracle Soundness]Let $\mathcal{R}$ be a residual relational constraint system. The evaluation oracle is a procedure yielding a judgment in the abstract domain $\mathbb{V}^\# = \{\mathsf{sat}, \mathsf{unsat}, \mathsf{unknown}\}$. We denote this evaluation as $\mathcal{R} \vdash_{oracle} \epsilon.valid \Downarrow v$. The oracle must be strictly sound with respect to the concretization space $\gamma_{\mathcal{R}}(\mathcal{R})$ such that proving a violation ($\mathsf{unsat}$) guarantees no valid concretizations exist:$$\mathcal{R} \vdash_{oracle} \epsilon.valid \Downarrow \mathsf{unsat} \implies \forall \mathcal{R}_c \in \gamma_{\mathcal{R}}(\mathcal{R}). \llbracket \mathcal{R}_c \rrbracket \downarrow \epsilon.valid = \mathsf{False} $$\end{axiom}\begin{definition}[Ideal Oracle]An oracle is considered \emph{ideal} if it is complete and never yields $\mathsf{unknown}$. For an ideal oracle, evaluation yields $\mathsf{sat}$ if and only if there exists at least one valid concretization:$$\mathcal{R} \vdash_{oracle} \epsilon.valid \Downarrow \mathsf{sat} \iff \exists \mathcal{R}_c \in \gamma_{\mathcal{R}}(\mathcal{R}). \llbracket \mathcal{R}_c \rrbracket \downarrow \epsilon.valid \neq \mathsf{False}$$\end{definition}

\subsection{Compiling Structure to Semantic Constraints}
\label{sec:narrowing}

We map the production rules and partial derivation trees to residual relational constraint systems. Then, we prove that the above holds through a state transition system. We begin with individual production mapping functions.
\begin{definition}[Attribute Grammar Production Mapping]
Let $p = (X_0 \to X_1 X_2 \dots X_n) \in P$ be a production rule in the grammar. The attribute grammar populates $E_p$ with functional equations for attributes $a$ defining parent synthesis $a \in \mathbf{S}(X_0)$ or child inheritance $a \in \mathbf{I}(X_i)$ at position $i \in \mathrm{N}$. Let $v_p = \{ x_0, x_1 \dots x_n \}$ be variables representing the production indices. The \emph{production mapping function} $\mathscr{P}_\pi\langle p \rangle$ is defined to convert local structural equations into absolute positional constraint rules with their sorts mapping to the attribute grammar domain:
$$\mathscr{P}_\pi \langle p \rangle = \left\{ \pi.i.a \rightarrow f(\pi.j.b), \dots) \;\middle|\; (x_i.a = f(x_j.b, \dots)) \in E_p
\land{} \text{sort}(f(x_j.b, \dots)) = D(a)
\right\}$$
Instantiating $p$ at address path $\pi$ generates rules through the coordinate mapping transformation $[x_0 \mapsto \pi] \cup [x_i \mapsto \pi \cdot i]$.
\end{definition}
Using our individual production mapping function, we define a global compilation function that works over any partial tree.

\begin{definition}[Global Compilation Function]
Let $t$ be a partial derivation tree. The global compilation function $\mathscr{C}_\pi\langle t \rangle$ maps an arbitrary partial derivation tree rooted at coordinate position $\pi$ to its complete residual relational constraint system through structural induction:
\begin{itemize}
    \item \emph{Base case}. If $t$ is an open hole $\Box_X \in \mathcal{H}$ of sort $X \in V$, compile to rules with a relational placeholder $\Box_X$ representing an unexpanded future boundary.
    $$\mathscr{C}_\pi \langle \Box_X \rangle = \bigcup_{a \in S(X)} \left\{ \pi.a \rightarrow \pi.\Box_X.a \right\}$$
    \item \emph{Inductive case}. If $t$ is a concrete structural node via production $p = (X_0 \to X_1 \dots X_n)$ and contains fresh holes representing subtrees $\Box_{X_1}, \dots, \Box_{X_n}$.
    $$\mathscr{C}_\pi \langle p(\Box_{X_1}, \dots, \Box_{X_n}) \rangle = \mathscr{P}_\pi \langle p \rangle \cup \bigcup_{i=1}^n \mathscr{C}_{\pi.i} \langle \Box_{X_i} \rangle$$
\end{itemize}
\end{definition}

\subsection{Core Equivalence and Soundness Meta-Theorems}
\label{sec:ag_theorems}
We now formally prove that our system is equivalent to a traditional evaluator.

\begin{lemma}
  The global compilation function $\mathscr{C}$ distributes homomorphically over structural refinement. For any partial syntax tree $t$ and valid substitution $\sigma = [\Box_X \leftarrow t']_\pi$:$$\mathscr{C}_\epsilon \langle t [\Box_X \leftarrow t']_\pi \rangle \equiv \mathscr{C}_\epsilon \langle t \rangle \cup \mathscr{C}_\pi \langle t' \rangle \setminus \mathscr{C}_\pi \langle \Box_X \rangle$$
\end{lemma}

\begin{theorem}[Equivalence of Declarative and Traditional Evaluation]
   Let $t_c \in T_{\Sigma_{\mathit{RAG}}}(\emptyset)$ be a complete, ground syntax tree. The satisfiability of its compiled constraint system under the structural choice variables $\Omega$ is strictly equivalent to traditional functional attribute evaluation.$$\llbracket t_c \rrbracket \downarrow \epsilon.\mathit{valid} = \mathsf{True} \iff  \mathscr{C}_\epsilon \langle t_c \rangle\vdash_\mathit{oracle}  \epsilon.\mathit{valid} \Downarrow \mathsf{sat}$$
\end{theorem}

\begin{lemma}[Monotonicity of Partial Constraints]
    For any partial syntax tree $t$ and any concrete tree $t_c \in \gamma(t)$ that refines it, the compiled constraints of $t$ safely over-approximate the semantics of $t_c$. If the partial constraint system is unsatisfiable, all concretizations are unsatisfiable:
    $$\mathscr{C}_\epsilon \langle t \rangle \vdash_\textit{oracle}  \epsilon.\mathit{valid} \Downarrow \mathsf{unsat} \implies \forall t_c \in \gamma(t). \; \llbracket t_c \rrbracket \downarrow \epsilon.\mathit{valid} = \mathsf{False}$$
\end{lemma}

\section{Semantic Constrained Decoding via Language Derivatives}
\label{sec:semantic_derivatives}
In this section, we lift context-free language derivatives to track the character-by-character construction of lexemes (Sec.~\ref{sec:lexical_abstraction}).
We define an inductive relational derivative operator over sequences of structural holes, modeling the progressive consumption of a string prefix (Sec.~\ref{sec:semantic_derivatives_operator}).
We conclude by demonstrating that our character-level derivative framework converges monotonically and guarantees both the soundness and completeness of real-time semantic pruning (Sec.~\ref{sec:monotonic_convergence}).

\subsection{Lexical Hole Semantics and Character Derivatives}
\label{sec:lexical_abstraction}

During constrained decoding, lexical tokens are not yielded atomically; they are constructed character by character. To evaluate semantic constraints speculatively over an in-progress token, we specialize the representation of terminal holes.
\begin{definition}[Lexical Holes]
Let $\Sigma_{\mathit{char}}$ be the lexical alphabet. During left-to-right decoding, terminal nodes are constructed incrementally.
Let $c \in \Sigma$ be a terminal category.
Let $\mathit{Pref}(\mathcal{L}(c))$ denote the set of all prefixes in $c$.
We define a \emph{Lexical Hole} as a stateful placeholder $\Box_{c \setminus w}$ where $c \in \Sigma$ is the target terminal category and $w \in \Sigma_{\mathit{char}}^*$ is the accumulated character prefix.A lexical hole is mathematically well-formed if and only if $w$ is a valid prefix of the token's language: $w \in \mathit{Pref}(\mathcal{L}(c))$. When the accumulated prefix fully satisfies the lexical definition ($w \in \mathcal{L}(c)$), the hole may be concretized into a solid terminal node $\blacksquare_{c \setminus w}$.
\end{definition}

An open lexical hole $\Box_{c \backslash w}$ speculatively accumulates characters as characters are consumed from left-to-right. Consumption terminates upon reaching a designated token boundary. If the completely generated lexeme string $w \in \Sigma_{\mathit{char}}^*$ legally satisfies the definition of the terminal category $c$ (denoted $w \in \mathcal{L}(c)$), the lexical hole may be transitioned to a concrete node.  We denote this finalized structural state using a solid terminal node $\blacksquare_{c \backslash w}$ representing the space of a single string $\{w\}$.
Syntactically, the filled-in placeholder directly equates to a standard singular leaf node within a concrete syntax tree: $$\blacksquare_{c \backslash w} \equiv w \in T_{\Sigma_{\mathit{RAG}}}(\emptyset)$$

The refinement of lexical holes is defined until the upper limit of some concrete lexical token:
$$\forall c \in \Sigma. \; \forall w \in \mathcal{L}(c). \; \forall w' \cdot x = w. \; \Box_{c \backslash \epsilon} \sqsubseteq \Box_{c \backslash w'} \sqsubseteq \Box_{c \backslash w} \sqsubseteq \blacksquare_{c \backslash w}$$

\subsection{Semantic Derivatives of Partial Trees}
\label{sec:semantic_derivatives_operator}

With the partial tree algebra and abstract lexical contexts defined, we now formally bridge the mechanics of predictive left-to-right parsing with our constraint framework. In standard formal language theory, the Brzozowski derivative yields the set of valid string suffixes given a prefix $x$~\cite{brzozowski_derivatives_1964}. This was later extended to Context-Free Languages through caching and fixed point solving~\cite{might_parsing_2011}. Our residual attribute grammar system was designed specifically with the caching mechanism in mind to offload context-free ambiguity onto relational fixed points.

First, we define a structural nullability operator over a placeholder sequence. We denote this using the subscript $\nu$. This operator checks if a sequence of holes can derive the empty string.\begin{definition}[Structural Nullability Operator]Let $\alpha \in (N \cup \Sigma)^*$ be a sequence of grammar symbols. The nullability operator $\Box_{\alpha \setminus \nu}$ statically evaluates whether the sequence can vanish into the empty string $\epsilon$:$$\Box_{\alpha \setminus \nu} = \begin{cases}
      \{\epsilon\} & \text{if } \alpha \xrightarrow{*} \epsilon \\
      \emptyset & \text{otherwise}
   \end{cases}$$For a single non-terminal hole $\Box_X$, $\Box_{X \setminus \nu} = \{\epsilon\}$ iff there exists a production $X \to \alpha$ such that $\Box_{\alpha \setminus \nu} = \{\epsilon\}$.\end{definition}

We define the derivative operator over a sequence hole $\Box_{\alpha \setminus x}$, representing the set of valid remaining symbol sequences after the prefix sequence $\alpha$ consumes the character $x \in \Sigma_{\mathit{char}}$.\begin{definition}[Derivative of a Sequence Hole]
Let $x \in \Sigma_{\mathit{char}}$ be the next consumed character. The derivative of an active lexical hole followed by a remaining syntactic sequence $\beta \in (N \cup \Sigma)^*$ is defined as:$$\Box_{c \setminus w}\beta \setminus x =
\begin{cases}
\emptyset & \text{if } wx \notin \mathit{Pref}(\mathcal{L}(c)) \\
\{ \Box_{c \setminus wx}\beta \} & \text{if } wx \in \mathit{Pref}(\mathcal{L}(c)) \land wx \notin \mathcal{L}(c) \\
\{ \Box_{c \setminus wx}\beta, \; \beta \} & \text{if } wx \in \mathcal{L}(c)
\end{cases}$$
For a single non-terminal hole $\Box_{X \setminus x}$, the derivative is the union of the derivatives of its productions $P_X$:$$\Box_{X \setminus x} = \bigcup_{(X \to \alpha) \in P_X} \Box_{\alpha \setminus x}$$\end{definition}

We introduce a control variable $\Omega_\pi \in P_X$ for every coordinate $\pi$ in the tree. This variable strictly binds the oracle to a single production choice for all attributes at that node. We introduce the guard operator $[\text{condition}] \triangleright \text{Rule}$, meaning the relational rule is only active if the condition is satisfied.\begin{definition}[Guarded Residual Compilation of Derivatives]Let $\Box_{X \setminus w}$ be a non-terminal hole of sort $X$ that has consumed prefix $w$. Let $P_X = \{p_1, \dots, p_k\}$ be the productions of $X$, where each $p_i = X \to \alpha_i$.We define a domain constraint for the position $\pi$, dictating that the structural choice variable $\Omega_\pi$ must map to exactly one valid production index:$$\mathcal{C}(\pi) = \left( \Omega_\pi \in \{1, \dots, k\} \right)$$The global compilation function $\mathscr{C}_\pi$ for the derivative hole maps to a set of guarded attribute equations:$$\mathscr{C}_\pi \langle \Box_{X \setminus w} \rangle = \mathcal{C}(\pi) \cup \bigcup_{i=1}^k \left( [\Omega_\pi = i] \triangleright \mathscr{C}_\pi \langle \Box_{\alpha_i \setminus w} \rangle \right)$$\end{definition}

\begin{definition}[Global Path Consistency]
Let $\vec{\Omega}$ be the set of all structural choice variables introduced in a derivative compilation $\mathscr{C}_\epsilon\langle\Box_{S \setminus w}\rangle$. The system must generate a global structural well-formedness constraint $\Phi_{struct}$ ensuring that if a parent node at coordinate $\pi$ selects production $p_i = X \rightarrow X_1 \dots X_n$, then the choice variables for any children $\pi.k$ are only active if $k \le n$.$$\Phi_{struct} = \bigwedge_{\pi} \left( (\Omega_\pi = i) \implies \bigwedge_{j=1}^{|p_i|} C(\pi.j) \right)$$The oracle's evaluation of the residual relational constraint system is strictly conjoined with $\Phi_{struct}$ to prevent the solver from satisfying semantic constraints via topologically disjoint derivations.
\end{definition}
\subsection{Meta-Theorems}
\label{sec:monotonic_convergence}

\begin{theorem}[Soundness of Relational Derivative]
Let $w \in \Sigma_{\mathit{char}}^*$ be a generated prefix, and $x \in \Sigma_{\mathit{char}}$ be the next consumed character. Let $\mathit{yield}(t_c)$ map to the flat string of a concrete tree $t_c$. If the constraint system produced by the derivative operator over the new prefix $wx$ is unsatisfiable, then no complete tree extending $wx$ exists in the semantic language:
$$
\mathscr{C}_\epsilon \langle \Box_{S \setminus wx} \rangle \vdash_\textit{oracle}  \epsilon.\mathit{valid}  \Downarrow \mathsf{unsat} \implies \forall t_c \in T_{\Sigma_{\mathit{RAG}}}(\emptyset). \; \left( wx \le \mathit{yield}(t_c) \right) \implies \llbracket t_c \rrbracket \downarrow \epsilon.\mathit{valid} = \mathsf{False}$$
\end{theorem}

\begin{theorem}[Completeness of Real-Time Convergence]
For any ground truth syntax tree $t_c$ that validly satisfies the top-level specification ($\llbracket t_c \rrbracket \downarrow \epsilon.\mathit{valid} = \mathsf{True}$), the declarative derivative system remains continuously satisfiable for every step of its yielded character sequence. For any prefix $w \le \mathit{yield}(t_c)$:
$$ \mathscr{C}_\epsilon \langle \Box_{S \setminus w} \rangle \vdash_\textit{oracle}  \epsilon.\mathit{valid}  \Downarrow \mathsf{sat}$$
\end{theorem}

\section{Symbolic Encoding and Relational Invariant Synthesis}
\label{sec:symbolic_encoding}
In this section, we bridge our abstract declarative semantics with an executable verification backend by compiling constraint lattices into first-order logical formulas.
We begin by translating our residual relational constraint systems into CHCs evaluated over a decidable background theory (Sec.~\ref{sec:chc}).
We then detail the inductive compilation of structural rules, flattening tree coordinates and semantic attributes into strict relational predicates (Sec.~\ref{sec:chc_compilation}).
Finally, we introduce logical cyclic aliasing to soundly memoize infinite derivative unrolling.
Ultimately, this encoding reduces the continuous semantic validation of partial generated prefixes to a standard reachability query in an SMT solver. (Sec.~\ref{sec:solving})

\subsection{Constrained Horn Clauses and Background Theories}
\label{sec:chc}
To operationalize the abstract evaluation oracle $\vdash_{\mathit{oracle}}$, we map our residual relational constraint systems CHCs over a first-order background theory $\mathcal{T}$.

\begin{definition}[Compilation to CHCs]
Let $\mathscr{R}$ be a set of uninterpreted relation symbols representing our semantic domains, structural choices $\Omega$, and partial tree attributes. A CHC is a closed first-order logic formula of the form $\forall \vec{x}.\; (\phi \land \bigwedge_{i=1}^n \mathit{Rel}_i(\vec{x}_i) \implies H(\vec{x}_h))$, where $\phi$ is a constraint formulated strictly within the decidable background theory $\mathcal{T}$ (e.g., linear integer arithmetic, bit-vectors, or algebraic datatypes), $\mathit{Rel}_i \in \mathscr{R}$ are uninterpreted relational queries defining the body, and the head $H$ is either a relational assertion in $\mathscr{R}$ or the contradiction symbol $\bot$.
\end{definition}

We map the residual relational constraint system's elements into flat relational predicates by turning  positional coordinates and semantic attributes into predicate arguments.
Let $X \in V$ be a non-terminal symbol whose associated attribute set $A(X)$ is partitioned into inherited attributes $\mathbf{I}(X) = \{i_1, \dots, i_m\}$ and synthesized attributes $\mathbf{S}(X) = \{s_1, \dots, s_n\}$. We map $X$ to a distinct uninterpreted relation $\mathit{Rel}_X \in \mathscr{R}$. The signature of this predicate is defined by sorting the tree position alongside the semantic domains of its attributes:$$\mathit{Rel}_X : \mathbb{N}^* \times D(i_1) \times \dots \times D(i_m) \times D(s_1) \times \dots \times D(s_n) \to \{\mathsf{True}, \mathsf{False} \} $$where $\mathbb{N}^*$ denotes the domain of tree addresses. The core insight of this translation is that positional addresses $\pi$ are treated as explicit symbols or sequence variables within the solver's background theory, rather than opaque memory references. This formulation allows the CHC engine to track the grammar's tree topology purely relationally.

\subsection{Inductive Rule Compilation}\label{sec:chc_compilation}

Compiling rules to CHCs is a relatively straightforward process which is split into terminal and non-terminal rules.

Terminal nodes $\Box_{c \setminus w}$, where $w \in \mathcal{L}(c)$, acts as our lexical axioms. Its synthesized attributes $\vec{a}_c$ are statically computable constants derived directly from the lexeme string $w$ and the node emits a base fact.
For a tree position $\pi$, this lexical constraint maps directly to the following categorical axiom:$$\forall \pi, \vec{a}_c.\; (\vec{a}_c = \text{eval}(w)) \implies \mathit{Rel}_c(\pi, \vec{a}_c)$$
For non-terminal expansions, the structural relational constraint rules in $\mathcal{R}$ translate into inductive Horn clauses that bind parent relations to the validity of their children which is strictly guarded by the
structural choice variable $\Omega_\pi$.

\begin{definition}[Non-Terminal Compilation]
Let $p_k = X_0 \to X_1 X_2$ be a binary production rule. The declarative validity of the parent predicate $\mathit{Rel}_{X_0}$ at coordinate $\pi$ logically depends on the relational satisfiability of both child predicates, the solver's structural assignment of production index $k$, and the conjunction of the local semantic attribute equations. $$\forall \pi, \vec{a}_0, \vec{a}_1, \vec{a}_2.\; \left[ \mathit{Rel}_{X_1}(\pi.1, \vec{a}_1) \land \mathit{Rel}_{X_2}(\pi.2, \vec{a}_2) \land (\Omega_\pi = k) \land \phi_{p_k}(\vec{a}_0, \vec{a}_1, \vec{a}_2) \right] \implies \mathit{Rel}_{X_0}(\pi, \vec{a}_0)$$
\end{definition}
Here, $\phi_{p_k}$ represents the background theory constraint over the defunctionalized attribute vectors, mechanically extracted from the original attribute grammar equations $E_{p_k}$. The relational compilation also naturally encodes any valid CFG topology, including left-recursive grammars

\subsubsection{Derivative Memoization} Taking the semantic derivative of recursive rules leads to infinite structural unrolling. In operational parsing, derivative-based frameworks mitigate this by memoizing or "caching" derivative states, compacting the infinite derivation tree into a finite cyclic graph. We map this cyclic topology directly into the CHC system.

\begin{definition}[Cached Derivative States]
 Let $\Box_{X \setminus w}^{\mathit{ref}(\pi')}$ represent a derivative state at current coordinate $\pi$ that has been structurally identified as identical to a previously encountered ancestor state at coordinate $\pi'$.
\end{definition}
The global compilation function $\mathscr{C}_\pi$ halts expansion and emits a relational alias. The current placeholder and the cached ancestor both evaluate the same semantic non-terminal $X$ over the same accumulated prefix $w$, so they belong to the same uninterpreted relation $\mathit{Rel}_X$. The compilation of the cached symbol simply binds the validity of the current position $\pi$ to the validity of the ancestor position $\pi'$ across all semantic attributes $\vec{a}$:$$\mathscr{C}_\pi \left\langle \Box_{X \setminus w}^{\mathit{ref}(\pi')} \right\rangle = \left\{ \forall \vec{a}.\; \mathit{Rel}_X(\pi', \vec{a}) \implies \mathit{Rel}_X(\pi, \vec{a}) \right\}$$
We inject an explicit logical cycle into the constraint system, so the "cached" constraints are directly and actively managed during string compilation.
\begin{lemma}[Soundness of Relational Memoization]
The injection of the cyclic logical alias $\{\forall \vec{a}. \mathit{Rel}_X(\pi', \vec{a}) \Rightarrow \mathit{Rel}_X(\pi, \vec{a})\}$ safely approximates the infinite derivative unrolling without introducing spurious least-fixed-point models in the background theory $\mathcal{T}$.
\end{lemma}

\subsection{Queries and Meta-Theorems}
\label{sec:solving}
Semantic validation is framed as a reachability problem that balance existential and universal quantification. Specifically, we must prove that there exists at least one semantically valid program completion (an existential search over the structural choices, $\vec{\Omega}$) such that for all evaluation steps along that path, no semantic errors occur (a universal safety guarantee). CHC rules are emitted by selecting specific structural completions governed by the $\Omega_\pi$ parameters. For a specific choice path $\vec{\Omega}$, we encode the top-level semantic specification (the requirement that the synthesized validity attribute at the root evaluates to true, $\epsilon.\mathit{valid} = \mathsf{True}$) as the primary goal clause of the CHC system. This query is formally expressed as a clause attempting to deduce a valid state from an invalid root state:$$\exists \vec{a}. \; \mathit{Rel}_S^{\vec{\Omega}}(\epsilon, \vec{a}) \land (\vec{a}_{\mathit{valid}} = \mathsf{False}) \implies \mathsf{False}$$
If the solver determines the system is satisfiable, proving that $\mathsf{False}$ is statically unreachable for at least one path of $\vec{\Omega}$, it guarantees that a valid structural completion of the prefix $w$ exists where no emitted CHC constraint evaluates to an invalid state. Consequently, the prefix remains semantically viable. Conversely, if the solver computes that $\mathsf{False}$ is reachable for all possible $\vec{\Omega}$ assignments, it demonstrates that no valid structural completion exists. The current prefix deterministically violates the grammar's attribute constraints, and the decoding engine soundly and immediately prunes this lexical path from the search space.

\paragraph{Unknown SAT result}
In the CHC pipeline, the solver may return $\texttt{sat}$, $\texttt{unsat}$, or $\texttt{unknown/timeout}$. The framework is only sound and complete relative to the underlying oracle, so in order to accommodate this, there are one of three execution modes: (1) \textbf{Ideal oracle mode}: sound and complete relative to a complete oracle. (2) \textbf{Conservative mode}: \texttt{unknown/timeout} $\to$ \texttt{reject}; sound but incomplete. (3) \textbf{Speculative mode}: \texttt{unknown/timeout} $\to$ \texttt{accept}; progress-oriented, later revalidated, not immediately sound.

We conclude this section by formally grounding the translation from our declarative relational systems to the background logic.\begin{theorem}[CHC Translation Soundness]Let $w$ be an accumulated token prefix, and let $\mathcal{K}$ denote the flat set of Constrained Horn Clauses generated by the translation of the partial tree. The logic system $\mathcal{K}$ possesses a valid least-fixed-point model over the background theory $\mathcal{T}$ if and only if the declarative derivative system $\mathscr{C}_\epsilon \langle \Box_{S \setminus w} \rangle$ is satisfiable.\end{theorem}

\section{Evaluation}
\label{sec:evaluation}

In this evaluation, we answer the following research questions:
\begin{enumerate}[label=RQ\arabic*),font=\bfseries]
    \item Does \textsc{AGDecode} effectively constrain large language model outputs against safety semantics?
    \item What is the computational cost of solving SMT constraints during generation?
\end{enumerate}

\subsection{Case Study: Safety Constraints for C}

In this case study, we use \textsc{AGDecode} to enforce specific safety properties when generating C programs.
Many previous type-safety works focus on TypeScript since it is a mainstream language susceptible to type errors in LLM generation \cite{mundlerTypeConstrainedCodeGeneration2025}.
Conversely, small to mid-sized LMs tasked with generating C code \cite{pearce_asleep_2022} are shown to include many semantic safety errors specific to the language,
motivating our choice to pick C.

To evaluate the expressivity of \textsc{AGDecode},  we examine a suite of C safety properties governed by path-sensitive data theories. These include the UBs: (1) null pointer dereferences, (2) division-by-zero, and (3) uninitialized variable accesses.
Consider the program snippet below which demonstrates all three of these erroneous conditions.
\begin{center}
{\small
\begin{verbatim}
void func(int a) {
    int *x, *y, *z; y = malloc(sizeof(int)); z = malloc(sizeof(int)); a = 0;
    if (y == NULL || a < 1) return; *x = *z; *y = 10 / a;
}
\end{verbatim}
}
\end{center}
To prove that this program is free of the specified undefined behaviors,
a framework must track values and path conditions. In this scenario, the assignment to the uninitialized pointer \verb|*x|, the dereference to \verb|*z| and
the division by the variable \verb|a| should not be allowed.

\paragraph{Mapping grammar rules}
For this case study, we map a large subset of C grammar to our framework, including memory logic (pointers, derefs, array subscripting). We focus on checking the properties defined above, thus we do not include a memory model. The semantic rules are modelled using standard verification rules. For example, a division expression maps from attribute grammars to a CHC as the following:
\begin{align*}
P_\mathit{Expr}(\Gamma_\mathit{inh}, \sigma_\mathit{inh}, \Gamma_\mathit{syn}, \sigma_\mathit{syn}, v)  \impliedby &
P_\mathit{Expr}^1(\Gamma_\mathit{inh}, \sigma_\mathit{inh}, \Gamma', \sigma', v_1) {}\land{}
P_\mathit{Expr}^2(\Gamma', \sigma', \Gamma_\mathit{syn}, \sigma_\mathit{syn}, v_2) {}\land{} \\&
v = v_1 \div v_2 \land{} v_2 \neq 0
\end{align*}

\subsection{Experimental Setup}
In this section, we detail the experimental methodology designed to evaluate the efficacy of \textsc{AGDecode}.

\paragraph{Implementation}
The implementation for \textsc{AGDecode} is written in Python. Our framework generates CHCs based on (1) a given input specification in a relational style, and (2) a generated prefix. Relational SMT constraints are compiled using a thunked graph representation in the \texttt{networkx} library.
Constraints are then outputted into the SMT-LIB v2 format and solved using the Z3 Spacer~\cite{de_moura_z3_2008, komuravelli_smt_2014} engine, which is highly optimized for infinite CHCs.

\paragraph{Benchmark and Tasks}
We evaluate \textsc{AGDecode} against the HumanEval C++ subset of the MultiPL-E dataset~\cite{10.1109/TSE.2023.3267446} containing 161 LLM code generation problems and corresponding C++ function declarations. To the best of our knowledge, no widespread LLM evaluation benchmark exists for C because the language lacks the standardized containers that test harnesses rely on to validate outputs. To accommodate, we mapped the C++ functions to equivalent C functions. For example:
\begin{align*}
\text{C++ original} &: \texttt{long sum\_squares(std::vector<long> lst)} \\
\text{C test case} &: \texttt{long sum\_squares(long* lst, unsigned lst\_size)}
\end{align*}
We limited the set of test cases to 77 that focus on memory and pointer accesses and ignored those with floating point types. Our prompt had basic instructions to ensure the output used the C grammar we support and that the output was memory-safe. We also added the instruction that all pointer arguments should be checked before being accessed.

\paragraph{Model, Parameters, Hardware}
Test cases are generated against the CodeLLaMA-13B~\cite{roziere2024codellamaopenfoundation} model across 3 temperatures (0.3, 0.5, 0.7). Our experiments are run on a AMD EPYC 9654 (Zen 4) 2.4 GHz CPU with 1 Nvidia H100 SXM for LLM generation.

\paragraph{Procedure}
For each test case, we generate a semantic-constrained LLM output while collecting the time required to verify each generation and the number of tries required per decoder token. The experiments will run in \emph{speculative} mode, so unknown or timeout queries are treated as successful verification. Speculative mode allows for decoder tokens to continue generating under the assumption that a future fast query would validate or invalidate the entire generated prefix. For each SMT query, we imposed a timeout of 5 seconds.

\subsection{Results}

\subsubsection{RQ1: Effectiveness}

\begin{figure}[t]
{\small \begin{Verbatim}
bool move_one_ball(long* arr, unsigned arr_size) {
    assert(arr);
    bool swapped = false;
    int pos = 0;
    long temp;
    while (!swapped && (pos < arr_size - 1)) {
        swapped = false;
        for (int j = 0; j < arr_size - 1 - pos ; ++j) {
            if (arr[
\end{Verbatim}
}
\caption{LLM excerpt output of the HumanEval\_109\_move\_one\_ball test case (temp=0.7) constrained in the semantic framework. Removing the \texttt{assert(arr)} line causes the monitor to disallow the final decoder token \texttt{[}.}
\label{fig:output_res}
\end{figure}

Examining the output shows that the conditions are able to be monitored by our elements, even across multiple paths. Figure~\ref{fig:output_res} demonstrates the prefix of an output that was verified. We verified by removing the first line \verb|assert(arr)|, which would cause the final decoder token emitted \verb|[| to fail generating.
Some scenarios tend to be harder than others to rectify safety concerns when generated at the decoder-token level, and this is largely dependent on the grammar of the language. For example, we saw that memory access using array subscripting \verb|arr[0]| could easily revert to  \verb|arr| and produce a simple check like \verb|arr == NULL|. However, pointer dereferencing \verb|*arr| in almost all cases lead to doomed generations as reverting to \verb|*| would require some dereferenceable variable to be generated after the fact.

\subsubsection{RQ2: Overhead}

\begin{table*}[t]
\centering
\caption{Latency distributions across valid and invalid generations.}
\label{tab:codellama_temp_latency}
\resizebox{\textwidth}{!}{
\begin{tabular}{l ccccc ccccc ccccc}
\toprule
& \multicolumn{5}{c}{\textbf{CodeLlama-13B ($T=0.3$)}} & \multicolumn{5}{c}{\textbf{CodeLlama-13B ($T=0.5$)}} & \multicolumn{5}{c}{\textbf{CodeLlama-13B ($T=0.7$)}} \\
\cmidrule(lr){2-6} \cmidrule(lr){7-11} \cmidrule(lr){12-16}
\textbf{Query Result} & $N$ & \% & Q1 (s) & Med (s) & Q3 (s) & $N$ & \% & Q1 (s) & Med (s) & Q3 (s) & $N$ & \% & Q1 (s) & Med (s) & Q3 (s) \\
\midrule
Invalid & 1,803 & 46.1\% & 0.001 & 0.001 & 0.153 & 1,580 & 43.9\% & 0.001 & 0.002 & 0.532 & 1,668 & 38.9\% & 0.001 & 0.001 & 0.133 \\
Valid & 1,196 & 30.6\% & 1.169 & 1.806 & 2.564 & 1,108 & 30.8\% & 1.419 & 1.950 & 2.479 & 1,692 & 39.4\% & 0.235 & 1.124 & 2.137 \\
Timeout & 912 & 23.3\% & --- & 5.000 & --- & 914 & 25.4\% & --- & 5.000 & --- & 931 & 21.7\% & --- & 5.000 & --- \\
\midrule
\textbf{{Total Checks}} & 3,911 & 100.0\% & --- & --- & --- & 3,602 & 100.0\% & --- & --- & --- & 4,291 & 100.0\% & --- & --- & --- \\
\bottomrule
\end{tabular}
}
\end{table*}

\begin{figure}[t]
\centering
\begin{tikzpicture}
\begin{axis}[
    width=\textwidth,
    height=5cm,
    xlabel={Query Index},
    ylabel={Latency (seconds)},
    title={Z3 Latency per Executed Query: HumanEval\_122\_add\_elements (temp=0.5)},
    legend columns=-1,
    legend style={/tikz/every even column/.append style={column sep=0.4cm}, font=\small},
    grid=major,
    grid style={dashed, gray!30},
    ymin=0, ymax=6.5,
    xmin=0, xmax=75,
    enlarge x limits=false,
]
\addplot[color=gray, opacity=0.3, mark=none, forget plot] coordinates {
    (1,0.7008) (2,0.6965) (3,5.0004) (4,1.7403) (5,1.8172) (6,0.1209) (7,0.3014) (8,0.1231) (9,0.1212) (10,0.1218) (11,5.0003) (12,2.3954) (13,0.1213) (14,0.1240) (15,0.3007) (16,0.4219) (17,0.0012) (18,0.1300) (19,0.1305) (20,0.4195) (21,0.1283) (22,0.4216) (23,0.3367) (24,0.3314) (25,0.0011) (26,0.1304) (27,0.3348) (28,2.1935) (29,1.9693) (30,1.9944) (31,0.3267) (32,0.1309) (33,0.3476) (34,0.1337) (35,0.4226) (36,0.1327) (37,0.0012) (38,0.1313) (39,5.0005) (40,2.0412) (41,0.1311) (42,0.1350) (43,1.3869) (44,0.0012) (45,1.3776) (46,1.6528) (47,0.0012) (48,0.8832) (49,4.6958) (50,5.0003) (51,0.9388) (52,0.9357) (53,0.1408) (54,0.1481) (55,0.0012) (56,0.1353) (57,0.1499) (58,0.1368) (59,0.1669) (60,0.1524) (61,0.2178) (62,0.1498) (63,0.0013) (64,0.1498) (65,0.2172) (66,0.1486) (67,0.1499) (68,0.7981) (69,5.0003) (70,5.0003) (71,2.0061) (72,1.5439) (73,1.5874) (74,2.7439) (75,1.4031) (76,2.8021) (77,2.0784) (78,1.4916) (79,1.5184) (80,1.5152) (81,2.0663) (82,2.8774) (83,5.0004) (84,1.5213) (85,1.5078) (86,2.1549) (87,0.9726) (88,5.0003) (89,1.4477) (90,0.0013) (91,1.3796) (92,1.3847) (93,1.3812)
};

\addplot[color=red, dashed, thick, domain=0:100, forget plot] {5.0};

\addplot[only marks, mark=*, color=orange, mark options={scale=0.8, fill=orange, opacity=0.7}] coordinates {
    (6,0.1209) (7,0.3014) (8,0.1231) (9,0.1212) (10,0.1218) (13,0.1213) (14,0.1240) (15,0.3007) (16,0.4219) (17,0.0012) (18,0.1300) (19,0.1305) (20,0.4195) (21,0.1283) (22,0.4216) (23,0.3367) (24,0.3314) (25,0.0011) (26,0.1304) (27,0.3348) (31,0.3267) (32,0.1309) (33,0.3476) (34,0.1337) (35,0.4226) (36,0.1327) (37,0.0012) (38,0.1313) (41,0.1311) (42,0.1350) (44,0.0012) (47,0.0012) (53,0.1408) (54,0.1481) (55,0.0012) (56,0.1353) (57,0.1499) (58,0.1368) (59,0.1669) (60,0.1524) (61,0.2178) (62,0.1498) (63,0.0013) (64,0.1498) (65,0.2172) (66,0.1486) (67,0.1499) (90,0.0013)
};
\addlegendentry{Invalid -- reject}

\addplot[only marks, mark=*, color=green!70!black, mark options={scale=1.2, fill=green!70!black, opacity=0.9}] coordinates {
    (1,0.7008) (2,0.6965) (4,1.7403) (5,1.8172) (12,2.3954) (28,2.1935) (29,1.9693) (30,1.9944) (40,2.0412) (43,1.3869) (45,1.3776) (46,1.6528) (48,0.8832) (49,4.6958) (51,0.9388) (52,0.9357) (68,0.7981) (71,2.0061) (72,1.5439) (73,1.5874) (74,2.7439) (75,1.4031) (76,2.8021) (77,2.0784) (78,1.4916) (79,1.5184) (80,1.5152) (81,2.0663) (82,2.8774) (84,1.5213) (85,1.5078) (86,2.1549) (87,0.9726) (89,1.4477) (91,1.3796) (92,1.3847) (93,1.3812)
};
\addlegendentry{Valid -- accept}

\addplot[only marks, mark=x, color=red, mark options={scale=2.0, thick}] coordinates {
    (3,5.0004) (11,5.0003) (39,5.0005) (50,5.0003) (69,5.0003) (70,5.0003) (83,5.0004) (88,5.0003)
};
\addlegendentry{Timeout -- speculative accept}

    \path[name path=bottom] (axis cs:\pgfkeysvalueof{/pgfplots/xmin}, \pgfkeysvalueof{/pgfplots/ymin}) -- (axis cs:\pgfkeysvalueof{/pgfplots/xmax}, \pgfkeysvalueof{/pgfplots/ymin});

    \path[name path=top] (axis cs:\pgfkeysvalueof{/pgfplots/xmin}, \pgfkeysvalueof{/pgfplots/ymax}) -- (axis cs:\pgfkeysvalueof{/pgfplots/xmax}, \pgfkeysvalueof{/pgfplots/ymax});

    \addplot [red!20] fill between [
        of=top and bottom,
        soft clip={domain=3:4}
    ];
    \addplot [red!20] fill between [
        of=top and bottom,
        soft clip={domain=11:12}
    ];
    \addplot [red!20] fill between [
        of=top and bottom,
        soft clip={domain=39:40}
    ];
    \addplot [red!20] fill between [
        of=top and bottom,
        soft clip={domain=50:51}
    ];
    \addplot [red!20] fill between [
        of=top and bottom,
        soft clip={domain=69:71}
    ];

\end{axis}
\end{tikzpicture}
\caption{Iterative verification latency during constrained decoding for the HumanEval\_122\_add\_elements example (temp=0.5). Valid and invalid represent cases when the entire prefix till generation was verified per query. Timeout represents a query that did not complete within 5 seconds, allowing an unverified token to generate. The red background represents the uncertain period between unverified and verified generations.}
\label{fig:micro_latency}
\end{figure}
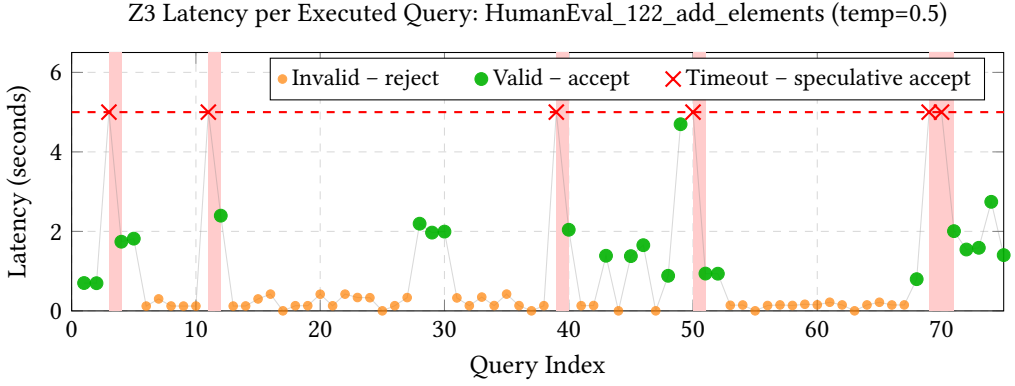

Table~\ref{tab:codellama_temp_latency} showcases the median per-decoder-token Z3 solving latency overhead across all three temperatures. About one-fourth (21.7\%-25.4\%) of queries reached the 5-second timeout. Of the remaining queries, the invalid queries were notably quick (0.532s Q3 at most, median 0.001s-0.002s), while valid queries had a larger range (0.235s Q1 at least to 2.564s Q3 at most). Surprisingly, the CHC queries were not as costly as expected. Spacer is well-optimized for infinite spaces, and performed especially well when proving invalid invariants of tokens.

\paragraph{Observation: Patterns of Verification Latency on Regular Boundaries} The realizability times per token varied wildly, and this typically came down to the structure of what was being queried, not the total length of the program.
Fig.~\ref{fig:micro_latency} showcases the query latency as generation progresses for one example, where every valid and timeout decoder token was accepted, increasing the prefix length being verified by successive queries. The timeout decoder tokens were speculatively accepted because a successive verification would verify the entire prefix. The red background indicates the time in which decoder-token generations are speculatively generated before being verified by a future query. This effect on query time is best demonstrated with an example. Consider the generation:
\begin{align*}
&\texttt{\{} &\;\to\;& \texttt{\{ arr[i} &\;\to\;& \texttt{\{ arr[i]; } \\
&\Box_{\mathit{Block}}  &\;\to\;& \textcolor{red}{\Box_{\mathit{Block} \backslash \texttt{arr[i}}} &\;\to\;& \textcolor{green!70!black}{\Box_{\mathit{Block} \backslash \texttt{arr[i];}}} \\
&\{ \Box_{\mathit{StmtList}} \} &\;\to\;& \{ \mathit{ArrIdx}(\mathit{arr}, \textcolor{red}{\Box_{\mathit{Expr} \backslash \texttt{i}}}); \,\ \Box_\mathit{\mathit{StmtList}}; \} &\;\to\;& \{ \mathit{ArrIdx}(\mathit{arr}, \textcolor{green!70!black}{i}); \,\ \Box_{StmtList}; \}
\end{align*}
where the first line is LLM output and the second and third lines are the corresponding internal constraint system states at the block level and statement list level.
The partially generated array index operator creates the placeholder $
\textcolor{red}{\Box_{\mathit{Expr} \backslash \texttt{i}}}$
an exploration for every possible expression that begins lexically with \texttt{i} (e.g., \texttt{i + ...}, \texttt{i = i * ...}, \texttt{ilist[...}) to index the array. This explosion of uniquely generated scenarios is challenging for naive solver and constraint encodings without efficient invariants, which we observed causing the majority of timeouts.
By the time that the full statement is realized, the hole is refined to a concrete expression $\textcolor{green!70!black}{i}$, which amounts to a simple CHC without recursive rules. Additionally, commonly used rules such as $\Box_{\mathit{StmtList}}$ are cached by an SMT solver. Scenarios like this highlight that verification time is not universally fast or slow as generation continues (i.e., a longer program is not necessarily slower to verify), but instead the effort depends on the possible valid program constructs. For this reason, while verification at the character or decoder-token level might not be feasible in every scenario, verification at statement boundaries can be.

Beyond constrained decoding, some modern LLM generation techniques allow multiple decoder tokens to be generated and backtracked if the complete generated code contains errors~\cite{li_rethinkmcts_2025}. Thus, an efficient and effective application of \textsc{AGDecode} consists of allowing tokens to be generated when verification queries time out, then backtracking if a later query invalidates previously allowed decoder tokens. If at any point, ambiguous decoder tokens were allowed and the later program verified, then no backtracking is required.

\subsection{Limitations}
While our approach provides strong theoretical guarantees, it is bounded by the provided grammar and specifications.  Manually defining comprehensive constraints is complex and highly susceptible to human error. Consequently, our evaluation was restricted to our chosen subset of C and representational ability, which notably omits a full memory model for solving time, and thus cannot enforce robust memory safety across unconstrained C programs. Furthermore, our limitation was severely limited by a lack of a backtracking implementation; in the cases where speculative token generation was later invalidated, any one of the tokens in-between could have been the cause of error, thus a valid implementation would have to consider which one to replace.

Verifying semantic constraints at runtime relies heavily on the underlying solver, which introduces issues of solver incompleteness. Because resolving deep relational queries can be computationally intractable which we remedy with a strict timeout policy. When a query times out, the system must either conservatively reject potentially valid tokens or accept unverified ones. This computational bottleneck results in severe token-level latency, rendering the approach mostly infeasible at the most fine-grained levels of generation.

Finally, our methodology is subject to the broader, systemic challenges of constrained decoding. Enforcing strict grammatical and semantic boundaries inherently distorts the language model's calibrated probability distribution. By masking out otherwise likely decoder tokens to satisfy the solver's constraints, the approach risks compromising the natural coherence and generative quality of the LLM output~\cite{ICLR2025_d98ce7a8}.

\subsection{Contrasting With A Coinductive Approach}

\begin{figure*}[t]
\centering
\resizebox{0.75\textwidth}{!}{%
\begin{tikzpicture}[
    >=stealth,
    node distance=1.5cm and 2cm,
    ast/.style={rectangle, draw, rounded corners, thick, minimum width=2.5cm, minimum height=0.8cm, align=center, fill=gray!10},
    state/.style={ellipse, draw=red!80, dashed, thick, fill=red!5, align=center, font=\small},
    attr/.style={rectangle, draw=blue!80, dotted, thick, fill=blue!5, align=center, font=\small},
    smt/.style={cylinder, shape border rotate=90, draw=black, aspect=0.25, thick, fill=gray!20, minimum height=1.5cm, minimum width=1.8cm, align=center},
    flow/.style={->, thick, blue!80, dashed},
    split/.style={draw, thick, dotted, gray}
]
\node[font=\bfseries] (title1) at (0, 1.5) {Coinductive};
\node[ast] (seq1) at (0, 0) {\texttt{Seq}};
\node[ast] (if1) at (-2, -1.8) {\texttt{If (x == NULL)}};
\node[ast] (deref1) at (2, -1.8) {\texttt{*x = 5}};
\draw[->, thick] (seq1) -- (if1);
\draw[->, thick] (seq1) -- (deref1);
\node[state] (s1) at (-2.5, -3.5) {$\Sigma_1: \{x = 0\}$};
\node[state] (s2) at (2, -3.5) {$\Sigma_2: \{x \neq 0\}$};
\node[state] (s3) at (3.5, -4.8) {$\Sigma_3: \{x = 1\}$};
\node[state] (s4) at (0.5, -4.8) {$\Sigma_4: \{x = 2\}$};
\node[font=\bfseries\Large, red] at (2, -5.5) {$\vdots$};
\node[font=\small, red, align=center] at (2, -6.2) {$\infty{}$};
\draw[->, red, dashed, thick] (if1) -- (s1);
\draw[->, red, dashed, thick] (deref1) -- (s2);
\draw[->, red, dashed, thick] (s2) -- (s3);
\draw[->, red, dashed, thick] (s2) -- (s4);
\draw[split] (5, 2) -- (5, -6.8);
\node[font=\bfseries] (title2) at (10, 1.5) {Inductive};
\node[ast] (seq2) at (10, 0) {\texttt{Seq}};
\node[ast] (if2) at (8, -1.8) {\texttt{If (x == NULL)}};
\node[ast] (deref2) at (12, -1.8) {\texttt{*x = 5}};
\draw[->, thick] (seq2) -- (if2);
\draw[->, thick] (seq2) -- (deref2);
\draw[flow] (seq2.west) to[bend right=30] node[left, font=\scriptsize] {$\text{in\_pc}$} (if2.north);
\draw[flow] (if2.east) to[bend right=30] node[right, font=\scriptsize, xshift=0.05cm] {$\text{out\_pc}$} (seq2.south);
\draw[flow] (seq2.east) to[bend left=30] node[right, font=\scriptsize] {$\text{out\_pc}$} (deref2.north);
\node[attr] (rel1) at (7.5, -4) {$\text{out\_pc} \iff$ \\ $\text{in\_pc} \land (x \neq 0)$};
\node[attr] (rel2) at (12.5, -4) {$\text{VC}: \text{out\_pc} \implies (x \neq 0)$};
\draw[->, dotted, thick, blue] (if2) -- (rel1);
\draw[->, dotted, thick, blue] (deref2) -- (rel2);
\node[smt] (smt) at (10, -6.2) {SMT \\ Oracle};
\draw[->, thick, -stealth] (rel1.south) |- (smt.west);
\draw[->, thick, -stealth] (rel2.south) |- (smt.east);
\end{tikzpicture}%
}
\caption{Comparison of constraint encodings for path-sensitive semantic decoding. \textbf{Left:} Coinductive approaches attempt to embed specific data environments (e.g., variable states) directly into the structural tree, breaking regularity and triggering an unbounded expansion of states. \textbf{Right:} \textsc{AGDecode} decouples the data domain from the structure using relational attributes, passing complex inductive VCs directly to an off-the-shelf SMT oracle.}
\label{fig:encoding_comparison}
\end{figure*}
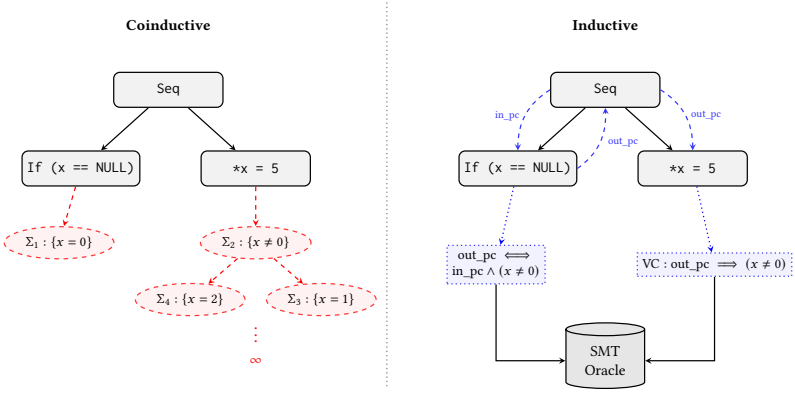

\textsc{ChopChop}'s coterm approach characterizes verification as the systematic pruning of non-well-founded program spaces. While this paradigm is elegant for specifying corecursive behaviours (e.g., liveness properties) and trace bisimulations, mapping inductive state-reachability properties (e.g., low-level, bit-precise C semantics) onto structural coterms introduces prohibitive representational overhead.
Coinduction relies on finding a greatest fixed point over a finite set of regular behaviours. For example, a coinductive tree attempting to model a loop would have to generate a completely unique structural state for every iteration, resulting in an infinite alphabet of symbols.

In essence, a coinductive pruner must simultaneously serve as both the operational semantic definition and the verification oracle. Because the pruner is formulated as a total function operating over a program space, complex spaces may need to be over- or under-approximated and any algorithmic decisions to do so must be explicitly baked into the pruner's structural traversal logic.

In contrast, \textsc{AGDecode} yields two primary structural advantages:

(1) \textbf{Global optimization context}. Rather than evaluating a lazy, step-by-step pruning function over an unfolded program space, our translation emits a system of relational constraints. This global view allows highly optimized SMT solving engines to exploit full contextual reasoning, leveraging techniques like conflict-driven clause learning and inductive generalization over the entire program state simultaneously.

(2) \textbf{Decoupled approximation mechanics}. Upon reaching a complex space, the choice to over- or under-approximate the valid program space is isolated to the oracle and not hard-coded into the front-end semantics.

Ultimately, our approach chooses incompleteness in the inductive invariant synthesis space to exploit highly optimized, automated data-theory reasoning. The coinductive approach makes the opposite trade-off: it secures completeness in behavioural structural properties at the expense of automated data-theory scalability.

Additionally, we do not provide a direct empirical comparison against \textsc{ChopChop} on the path-sensitive safety properties of our C case study. These properties are defined over reachable states distinguished by bit-precise values, and as argued above such states have no compact coterm representation. Expressing them in \textsc{ChopChop}'s coinductive setting requires a distinct structural symbol for each valuation, so the state space grows too large to make evaluation on this benchmark practical.

\section{Related work}
\label{sec:relatedwork}
\textbf{Constrained Decoding}
enforces structural, syntactic, or semantic constraints at generation time by masking invalid tokens given some rules~\cite{nagyChopChopProgrammableFramework2025, park_grammar-aligned_2025}.

\textbf{Syntactic constrained decoding. }
The most straightforward constrained decoding methods act at the  structural and syntactic levels \cite{nagyChopChopProgrammableFramework2025, beurer-kellner_guiding_2024, park_grammar-aligned_2025, geng_grammar_2023, willard_efficient_2023, scholak_picard_2021}. These approaches typically consume a valid grammar and build a parser that advances as tokens are consumed. At each step, the valid next tokens are found through a search of possible constructions in the grammar given a prefix. Syntactic constrained decoding, however, does not constrain on requirements that involve context, complex rules, meaning, or any other semantic constraints. \textsc{AGDecode} goes beyond the syntactic guarantees these methods provide.

\textbf{Semantic constrained decoding},
such as type-level constrained decoding or correctness guarantees, are much more complex and time consuming \cite{nagyChopChopProgrammableFramework2025, mundlerTypeConstrainedCodeGeneration2025, poesia_synchromesh_2022, agrawal_monitor-guided_2023}. Although both semantic and syntactic constrained decoding methods constrain the possible set of next tokens based on some prefix, semantic approaches go much further. Instead of maintaining a basic parse state, semantic approaches check whether a prefix can complete into a program satisfying some property. \citet{mundlerTypeConstrainedCodeGeneration2025} introduce type-constrained decoding. Their approach addresses the challenge of well-typed coding. This approach maintains a partial abstract syntax tree annotated with type-relevant context, and soundly ensures that all continuations adhere to this context. Although this approach soundly ensures well-typedness, they are limited to a single semantic property. More generally, \citet{nagyChopChopProgrammableFramework2025} open semantic constrained decoding to manual constraining. This approach lifts the analysis to a concise representation of all possible abstract trees, and, given a manually-written tree traverser, prunes these trees to find the set of valid continuations. Constraints must thus be encoded as the pruners themselves, strongly coupling the checking and the semantics, and introducing significant representation overhead. \textsc{AGDecode} completely separates these two by delegating the checking to external solvers, such as off-the-shelf SMT solvers, and allows for constraints to be written clearly and concisely in first order logic.

\textbf{Formal Verification}
approaches provide correctness guarantees by combining LLM output with machine-checkable proofs or symbolic analysis \cite{councilman_towards_2025, anshumaan_llm-based_2025, bhatia_verified_2024}. Unlike constrained decoding, formal verification approaches typically act post-hoc, requiring complete output, and often require generation iterations with verifier feedback. Instead, we apply formal methods on partial generations to enforce constraint satisfaction \emph{during} the token generation process.

\textbf{Repair }
 methods focus on post-hoc fixes through error analysis.
This can be execution-based analysis, such as by evaluating execution traces and unit test outputs \cite{chen_teaching_2023}. A parallel line of work is Automated Program Repair (APR), where LLMs iteratively repair and explore the repair space \cite{xia_less_2022, zhang_coder_2022}. Repair methods act post-hoc on already-generated code, whereas \textsc{AGDecode} prevents semantically invalid output from being generated in the first place.

\textbf{Search.}
Search-based methods improve correctness by generating multiple candidate programs and selecting or combining the best, rather than constraining or repairing a single output \cite{zhang_coder_2022, wang_mathbbuscd_2024, li_rethinkmcts_2025}.
Search-based methods improve correctness by sampling and selecting among multiple outputs, but do not enforce semantic validity during generation as our pruning approach does.

\section{Conclusion}
\label{sec:conclusion}

We present \textsc{AGDecode}, a programmable framework for semantic constrained decoding based on realizability over partially constructed syntax trees. By compiling semantic specifications into relational constraints solved by SMT and CHC backends, \textsc{AGDecode} supports rich inductive program properties while cleanly separating language semantics from verification machinery. We prove that \textsc{AGDecode} is sound and complete relative to an ideal verification oracle. Our experimental evaluation demonstrates that it can effectively enforce path-sensitive safety properties during C program generation. The results demonstrate feasibility but also show that token-level verification has significant tail latency; this motivates batching and speculative validation.

\paragraph{Future Work}
We plan to explore the integration of the SMT oracle to actively guide an LLM's next token choice in order to increase constrained decoding output. Additionally, we plan to explore the integration of backtracking to guide refined iterative generation of \emph{sections} of LLM outputs.

\bibliography{bibfile}

\end{document}